\documentclass{aa}  

\usepackage{graphicx}
\usepackage{txfonts}
\usepackage{xcolor}
\usepackage{ulem}
\begin{document}

   \title{Mapping Luminous Hot Stars in the Galaxy}

   \author{E. Zari
          \inst{1},
           H.-W. Rix \inst{1},
           N. Frankel \inst{1},
           M. Xiang \inst{1},
           E. Poggio \inst{2,3},
           R. Drimmel \inst{2},
           and A. Tkachenko \inst{4}
         }

   \institute{Max-Planck-Institut f\"ur Astronomie, K\"onigstuhl 17 D-69117 Heidelberg, Germany \\
    \and
    Osservatorio Astrofisico di Torino, Istituto Nazionale di Astrofisica (INAF), I-10025 Pino Torinese, Italy  \\
    \and
    Universit\'e C\^ote d’Azur, Observatoire de la C\^ote d’Azur, CNRS, Laboratoire Lagrange, France \\
    \and
    Institute of Astronomy, KU Leuven, Celestijnenlaan 200D, 3001, Leuven, Belgium}

   \date{Received xxx; accepted yyy}

 
  \abstract
  {Luminous hot stars ($M_{K_s} \lesssim 0 \, \mathrm{mag}$ and $T_{\mathrm{eff}}\gtrsim 8000$ K) dominate the stellar energy input to the interstellar medium (ISM) throughout cosmological time, they are laboratories to test theories of stellar evolution and multiplicity, and they serve as luminous tracers of star formation in the Milky Way and other galaxies. 
  Massive stars occupy well-defined loci in colour-colour and colour-magnitude spaces, enabling selection based on the combination of \textit{Gaia} EDR3 astrometry and photometry and 2MASS photometry, even in the presence of substantive dust extinction.
  In this paper we devise an all-sky sample of such luminous OBA-type stars, designed to be quite complete rather than very pure, to serve as targets for spectroscopic follow-up with the SDSS-V survey.   To estimate the purity and completeness of our catalogue, we derive stellar parameters for the stars in common with LAMOST DR6 and we compare the sample to other O and B-type star catalogues. We estimate "astro-kinematic" distances  by combining parallaxes and proper motions with a model for the expected velocity and density distribution of young stars; we show that this adds useful contraints on the stars' distances, and hence luminosities.  
  With these distances we map the spatial distribution of a more stringently selected sub-sample across the Galactic disk, and find it to be highly structured, with distinct over- and under-densities. The most evident over-densities can be associated with the presumed spiral arms of the Milky Way, in particular the Sagittarius-Carina and Scutum-Centaurus arms.  Yet, the spatial picture of the Milky Way's young disc structure emerging in this study is complex, and suggests that most young stars in our Galaxy ($t_{age}<t_{dyn}$) are not neatly organised into distinct spiral arms.
  The combination of the comprehensive spectroscopy to come from SDSS-V (yielding velocities, ages, etc..) with future \textit{Gaia} data releases will be crucial to reveal the dynamical nature of the spiral arm themselves.
  
}
 


    \authorrunning{Zari et. al}
   \maketitle
%

\section{Introduction}
Luminous and hot stars are massive, and hence rare and short lived. However, they play decisive roles across different fields of astrophysics.

They dominate the interaction between stars and the interstellar gas and dust (ISM) in their host galaxies, by heating or ionising those components; those stars also interact with the ISM through powerful stellar winds and eventually supernova explosions \citep{MacLow2004, Hopkins2014}.  Luminous and hot stars must indeed have played  an  important  role  in  galaxy  evolution throughout cosmic time, via their intense winds, ultraviolet radiation fields, chemical processing, and explosions \citep{Haiman1997, Douglas2010, Bouwens2011}.

Luminous and hot stars are inevitably young, and therefore they can serve as tracers of recent massive star formation. Although they make up an insignificant fraction of the overall stellar mass, they contribute a major portion of the light of the disc and they can thus probe the spiral structure and young disc kinematics of our and other galaxies \citep{Xu2018, Chen2019, Dobbs2014, Kendall2011, Kendall2015}.

They are, directly or through their remnants, decisive drivers of the chemical evolution for many elements in the periodic table and they serve as crucial laboratories for stellar evolution in this arguably least well-understood regime of stellar physics. 

Massive stars are born  predominantly as members of binary and  multiple  systems  \citep{Sana2012, Sana2014, Kobulnicky2014, Moe2017}. As a consequence, most of them are expected to undergo strong binary interaction, which drastically alters their evolution \citep{Podsiadlowski1992, VamBever2000, O'Shaughnessy2008, deMink2013, Langer2020}. 

Finally, massive (binary) stars are the only channel to yield binaries that involve black holes and neutron stars in the disc of the Milky Way. Therefore, understanding massive star binaries, also as they evolve away from their zero-age main sequence, is indispensable for understanding the distribution of gravitational wave events. 
For all the above reasons, a multi-epoch, spectroscopic census of luminous and hot stars across the Galaxy, providing spatial and dynamic information together with estimates for  masses, ages, metallicity, multiplicity, and other spectroscopic information is needed. The SDSS-V survey \citep[][Kollmeier, J., Rix, H.-W., et al., in prep]{Kollmeier2017, Bowen1973, Gunn2006, Smee2013, Wilson2019} will provide such a comprehensive, multi-epoch spectroscopic program on hot and massive stars, which -- however -- must be based on a clear and quantitative selection function to enable rigorous subsequent population studies.

There is no universal precise 'definition' across different sub-communities within astrophysics of when a star is 'luminous', 'hot', and 'massive'. 'Hot' may either mean that its $T_{\mathrm{eff}}$ is sufficiently high that the spectrum is dominated by H and He lines, rather than metal lines (i.e. OBA stars); or it may mean that $T_{\mathrm{eff}}$ is sufficiently high that the star produces significant amounts of ionising radiation (O and early B stars). 'Massive' may either mean, massive enough to 'go Supernova' or 'to have a convective core'.

For the current context we choose an operative definition of {\sl luminous and hot} as $T_{\mathrm{eff}} \gtrsim 8000$ K and $M_K\lesssim 0$ mag, or -- loosely speaking -- OBA stars.
This is the set of stars that addresses the science sketched above. Of course some aspects, such as ISM ionisation, are only affected by O and early B stars; but it is important to remember that even the most massive stars are not ionisingly hot during all of their evolution. Therefore we aim in our sample definition for a sample that is hot enough to eliminate all the luminous red giant branch (RGB) and  asymptotic giant branch (AGB) stars, but include most evolved phases of massive stars. Stars with $T_{\mathrm{eff}}\ge 8000$K can be photometrically pre-selected across the entire sky; yet efficient discrimination of 10,000~K from 20,000~K by means of photometry only would be nearly impossible in the presence of dust.

In light of the science goals above, in this paper we present the selection of the target sample of massive stars of the SDSS-V survey. We describe an approach to use kinematics to improve distance estimates. Finally we use a "clean" sub-set of the target sample to study the structure of the Milky Way disc as traced by young, massive stars.

The selection of our target sample is based on the combination of \textit{Gaia} EDR3 \citep{Prusti2016, Brown2020} astrometry and photometry and 2MASS \citep{Skrutskie2006} photometry. Similar samples of stars in the Upper Main Sequence (UMS) were already selected by \cite{Poggio2018} and \cite{Romero2019} to characterise the structure and the kinematic properties of the Galactic warp. \cite{Poggio2018} used a combination of \textit{Gaia} DR2 and 2MASS colours to select stars of spectral type earlier than B3, and obtained a sample of 599 494 stars. 
\cite{Romero2019} used only \textit{Gaia} DR2 photometry to select stars brighter than $M_{G, 0}  = 2 \, \mathrm{mag}$ (thus likely including also A-type stars), and obtained a sample of 1 860 651 stars.  Our sample definition (described in Section \ref{sec:data}) lies somewhat in the middle between those by \cite{Poggio2018} and \cite{Romero2019}.  In Section \ref{sec:dist_estimates} we describe the method that we use to estimate distances.
In Section \ref{sec:filtered_sample} we clean the target sample for sources with spurious astrometry and we filter out older contaminants. In Section \ref{sec:results} we use this filtered sample to study the 3D space distribution of the sources and we compare the distribution of OBA stars with other tracers of massive star formation and spiral arm structure.
In Section \ref{sec:discussion} we discuss our findings in the context of the structure and nature of the spiral arms of the Milky Way. Finally we summarise our results and draw our conclusions in Section \ref{sec:conclusion}.

\section{Target sample}\label{sec:data}
In this Section we 
present the criteria that we applied to select the target sample for the SDSS-V spectroscopic survey, and we describe its characteristics in terms of sky distribution, magnitude and luminosity distribution, variability, purity, and completeness.  

\subsection{Astro-photometric selection} \label{sec:selection}
To select our sample, we use  \textit{Gaia} EDR3 photometry \citep{Riello2020} and astrometry \citep{Lindegren2020} combined with 2MASS photometry. The cross-match betweeen \textit{Gaia} and 2MASS is not yet provided in the \textit{Gaia} archive (Marrese et al., 2021, in prep) \footnote{\url{https://archives.esac.esa.int/gaia}}. For this reason we performed first a cross-match with \textit{Gaia} DR2, then a cross-match with 2MASS. The ADQL query for this cross-match is provided in Appendix \ref{sec:query}.

We restrict our query to the sources with $G < 16 \, \mathrm{mag}$. The motivation of this condition is twofold.  On the one hand, it is a SDSS-V technical requirement: at $G = 16$ mag we can obtain a S/N = 75 with 15 minute exposures with the BOSS spectrograph \citep{Smee2013}. On the other hand, for  $G > 16 \, \mathrm{mag}$ the \textit{Gaia} EDR3 parallax precision rapidly deteriorates \citep[see e.g. Table 3 in][]{Brown2020},
thus a) estimating absolute magnitudes (which are crucial to our selection) becomes non-trivial and b) inferring distances strongly depends on the choice of the prior (see Appendix \ref{sec:kin_dist} and Section \ref{sec:dist_estimates}).  
    
We define a proxy for the absolute magnitude of a star in the $K_s$ band, $\Tilde{M}_{K_s}$, 
    and we require $\Tilde{M}_{K_s} < 0$. This translates to the following parallax condition:
    \begin{equation}\label{eq:MK0}
    \varpi < 10^{(10 - K_s - 0.0)/5},
    \end{equation}
and aims at selecting a reasonably-sized sample of bright stars. 
The value $M_{K_s} = 0 \, \mathrm{mag}$ corresponds roughly to a B7V-type star \citep{Pecaut2013}. To include also later B-type stars we should require $\tilde{M}_{K_s} < 1 \, \mathrm{mag}$, as $M_{K_s} = 1 \, \mathrm{mag}$ roughly corresponds to the absolute magnitude of a A0V-type star. We chose however $\tilde{M}_{K_s} = 0 \, \mathrm{mag}$ as a good compromise between completeness and contamination (see Section \ref{sec:completeness}).
 
We stress that for the survey target list we do not require any condition on the photometric quality of 2MASS, nor on the photometric errors,  nor on the parallax accuracy or the goodness of fit of the astrometric solution. This is motivated by the fact that we are aiming to obtain a known, well defined selection function: not only measurement errors do not describe intrinsic physical properties of the sources, and thus should not enter the selection criteria, but they are also difficult to model. 

At this point our sample, shown in the $(J - K_s)_0$ vs $G_{\mathrm{BP}} - G_{\mathrm{RP}} $ colour-magnitude diagram in Fig. \ref{fig:cmd_ell}, consists of luminous stars, which could either be OBA, RGB, and AGB stars.  OBA stars fall in the region highlighted by the black ellipse. The stars in the horizontal stripes are stars with large 2MASS photometric errors or 2MASS photometric quality flags (\texttt{ph\_flag}) different than "AAA".

\begin{figure*}
    \centering
    \includegraphics[width = 0.45\hsize]{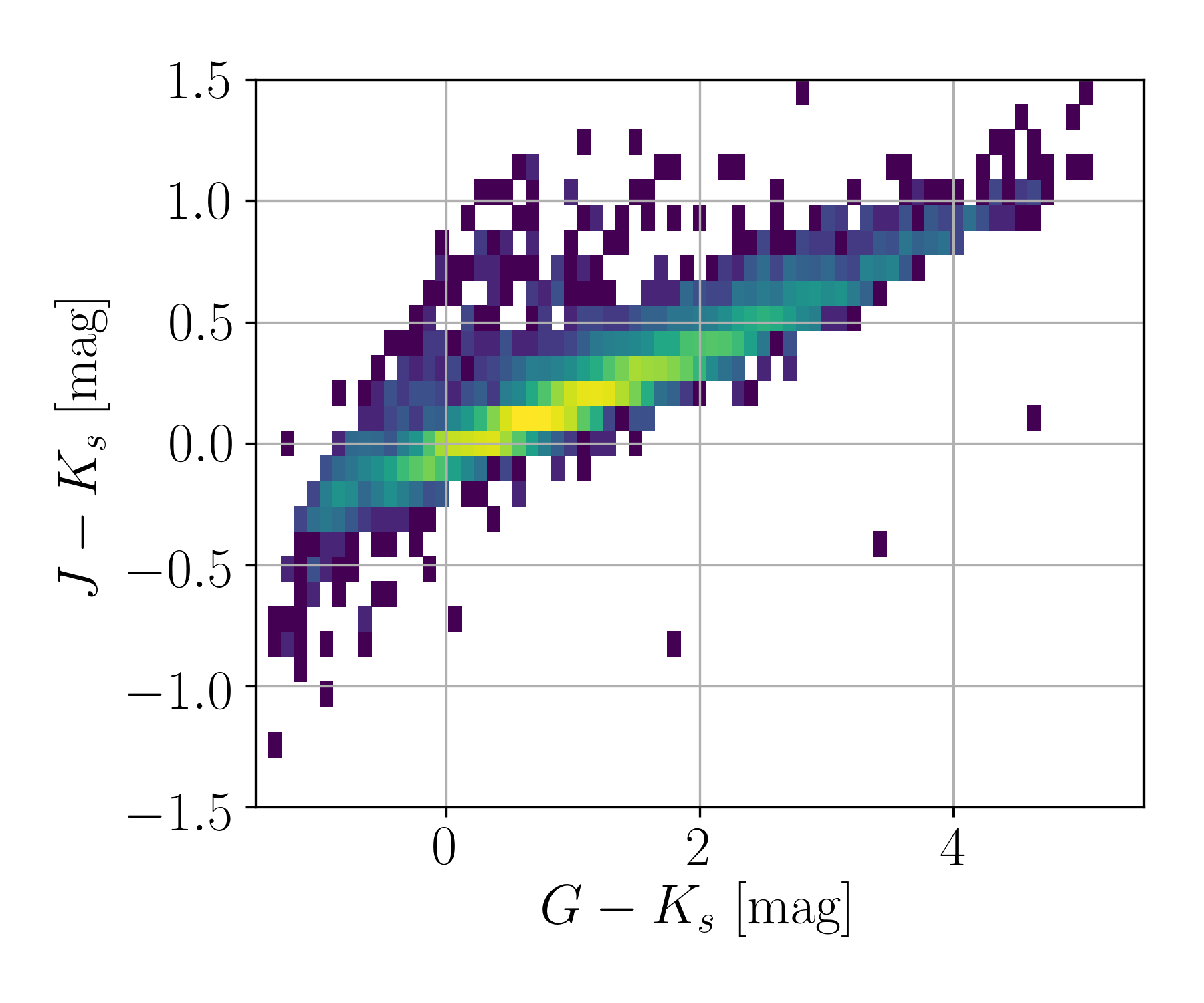}
    \includegraphics[width = 0.45\hsize]{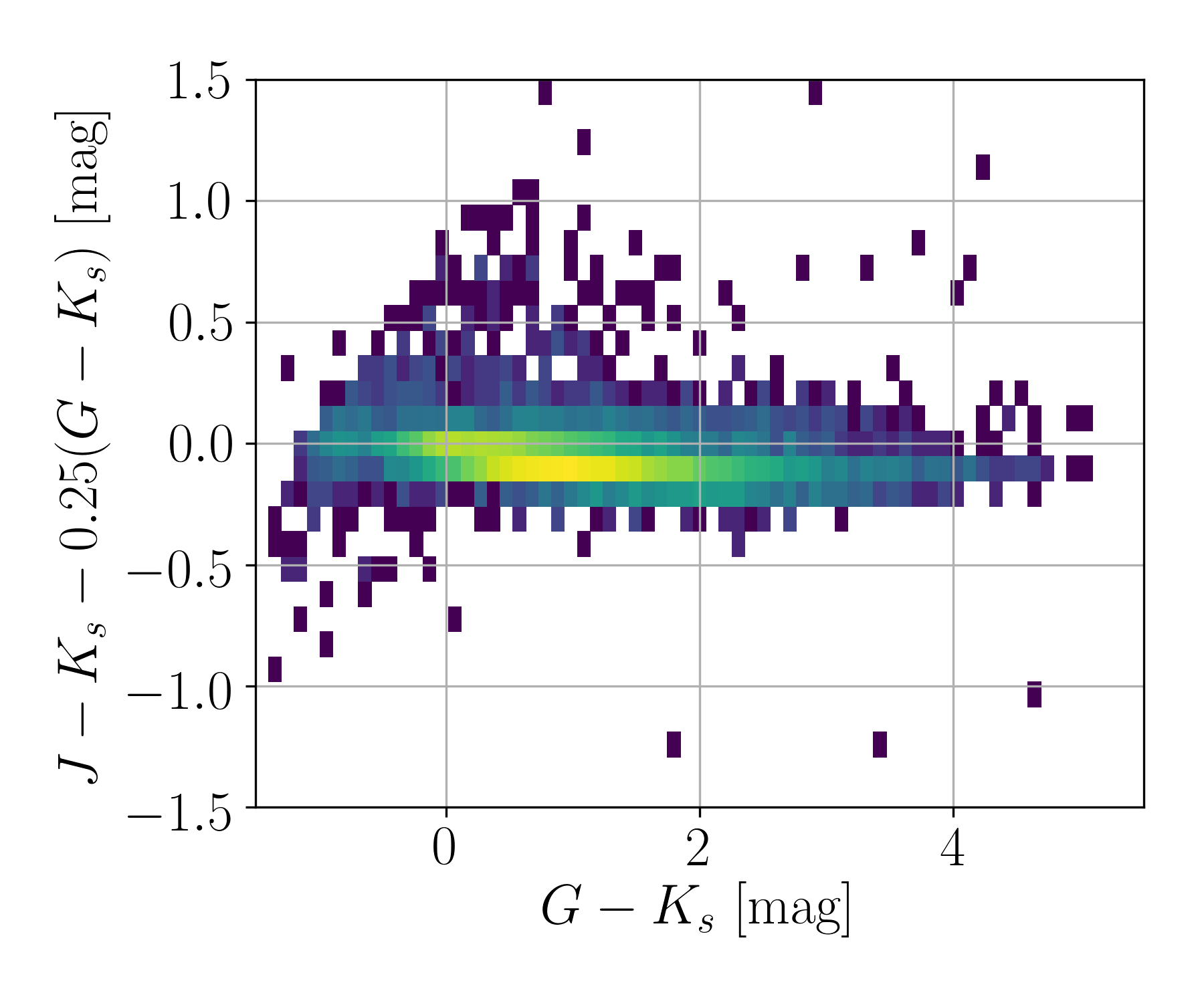}
    \caption{2D histogram of $J-K_s$ vs. $G- K_s$ (left) and $(J - K_s)_0$ and $G - K_s$ (right) for the stars in the LAMOST sample by \cite{Liu2019} (see main text for more details on the sample).}
    \label{fig:appendix1}
\end{figure*}

To select OBA stars we use simple photometric cuts, roughly following the procedure outlined in
\cite{Poggio2018} (see in particular their Fig. 1).
The first step of our selection consists in selecting  sources with:  
\begin{equation}\label{eq:3}
(J - K_s)_0 < 0.1 \, \mathrm{mag} \hspace{0.1cm} \mathrm{and} \hspace{0.1cm} (J - K_s)_0 > -0.3 \, \mathrm{mag}.
\end{equation}
The first condition excludes most of the red sources, the second condition removes  sources with unnaturally blue colours.
We obtained the $(J - K_s)_0$ colour by applying the following Equation:
\begin{equation}\label{eq:correction}
    (J - K_s)_0 = (J - K_s) - 0.25(G - K_s),
\end{equation}
which we derived by noticing that, for the spectroscopically confirmed O and B-type stars selected by \cite{Liu2019}, the $(J - K_s)$ and $(G - K_s)$ colours are linearly correlated (see Fig. \ref{fig:appendix1}, left), and by assuming that the difference between the $J$ and $K_s$ magnitudes for such stars is close to zero \citep[see e.g.][]{Pecaut2013}.

\begin{figure}
    \centering
    \includegraphics[width = \hsize]{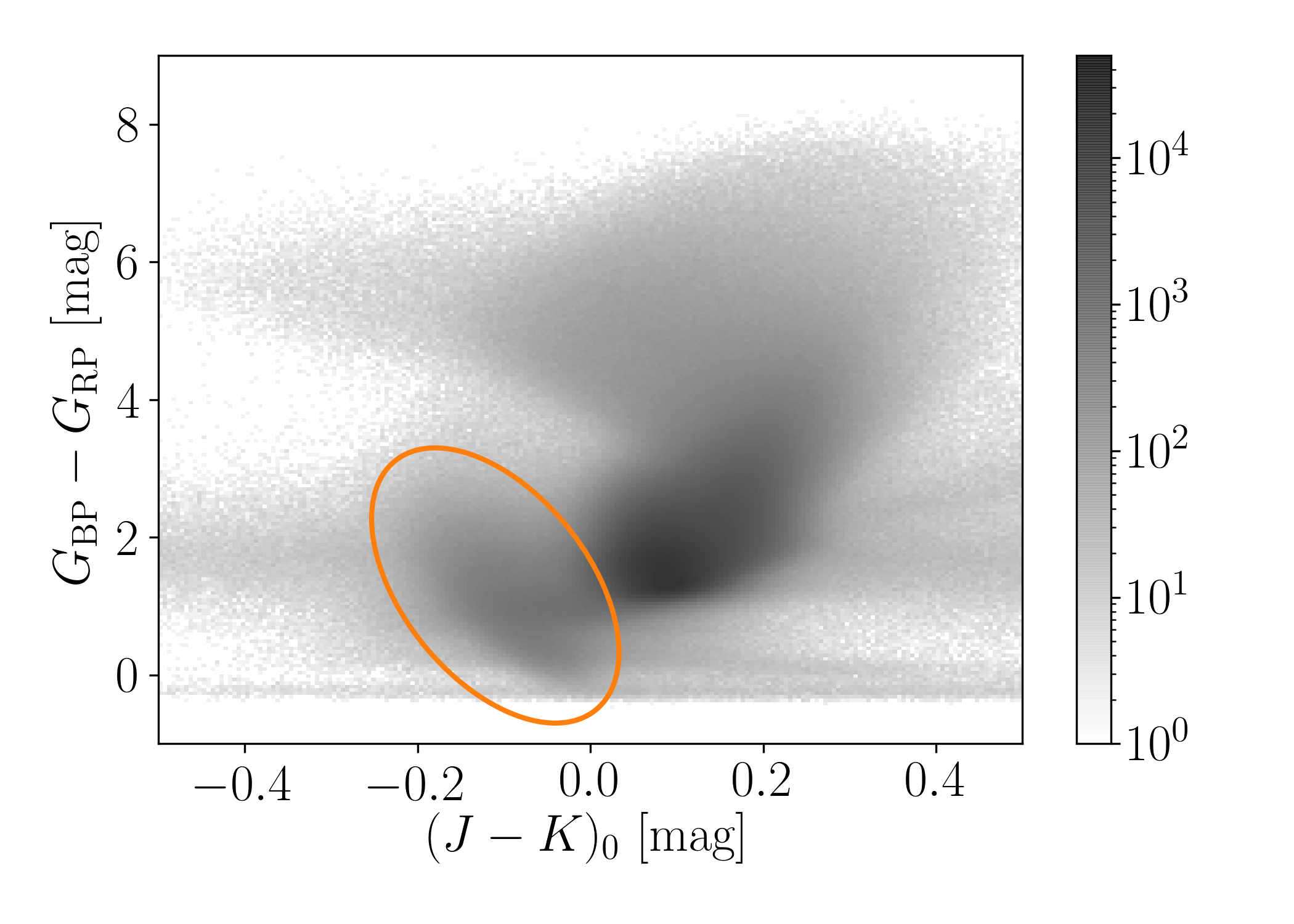}
    \caption{$(J - K_s)_0$ vs. $G_{\mathrm{BP}}  - G_{\mathrm{RP}}$ colour-magnitude diagram. OBA stars roughly fall in the region highlighted by the orange ellipse. This representation shows that hot stars stand out qualitatively in colour-colour space.}
    \label{fig:cmd_ell}
\end{figure}
\begin{figure}
    \centering
    \includegraphics[width = \hsize]{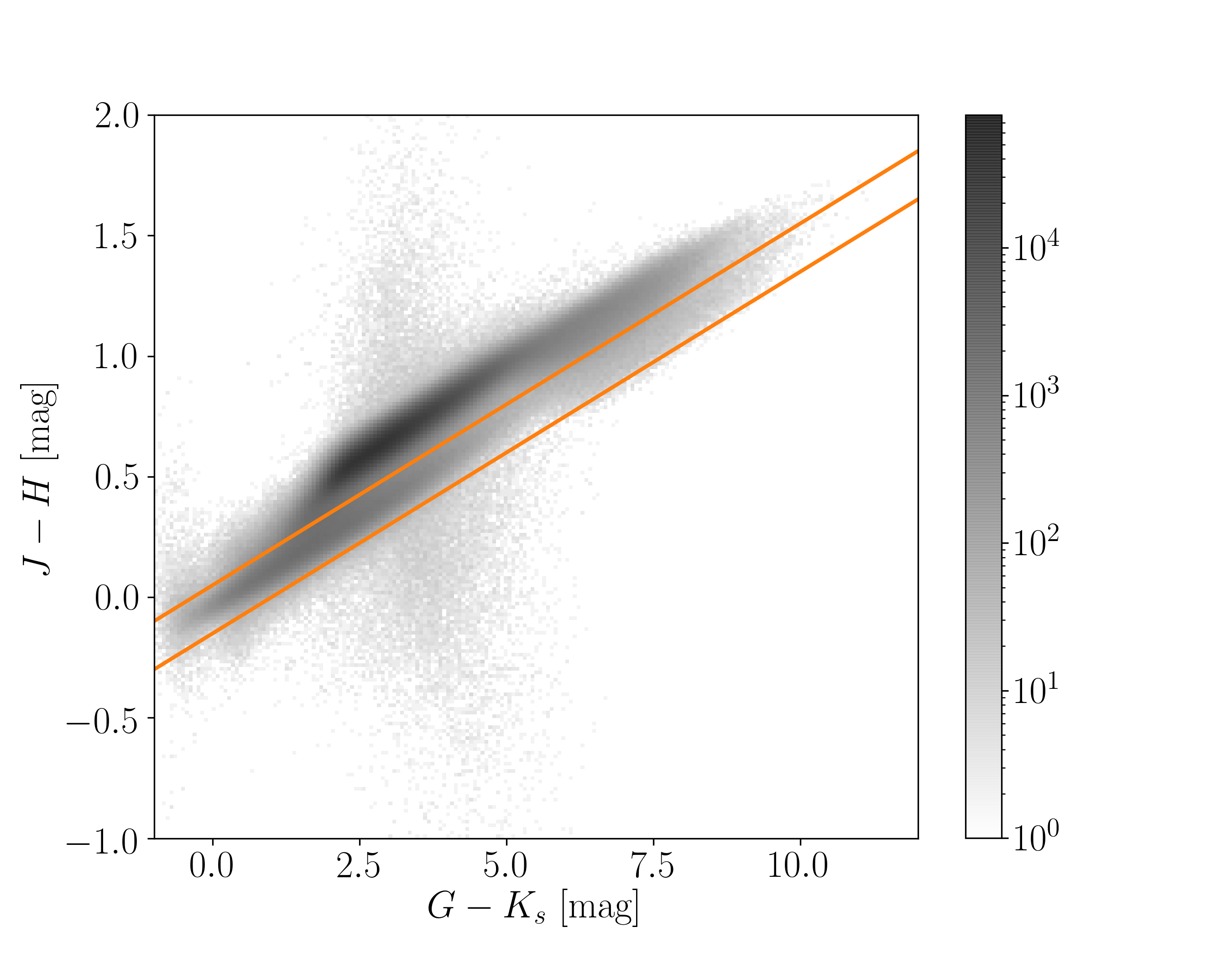}
    \caption{$J-H$ vs. $G - K_s$ colour-colour diagram of the sources with $(J - K_s)_0 < 0.1 \, \mathrm{mag}$ and $(J - K_s)_0 > -0.3 \, \mathrm{mag}$. The solid lines delineate the criteria for selecting OBA stars and are defined in the text.}
    \label{fig:cmd2}
\end{figure}
\begin{figure}
    \centering
    \includegraphics[width = \hsize]{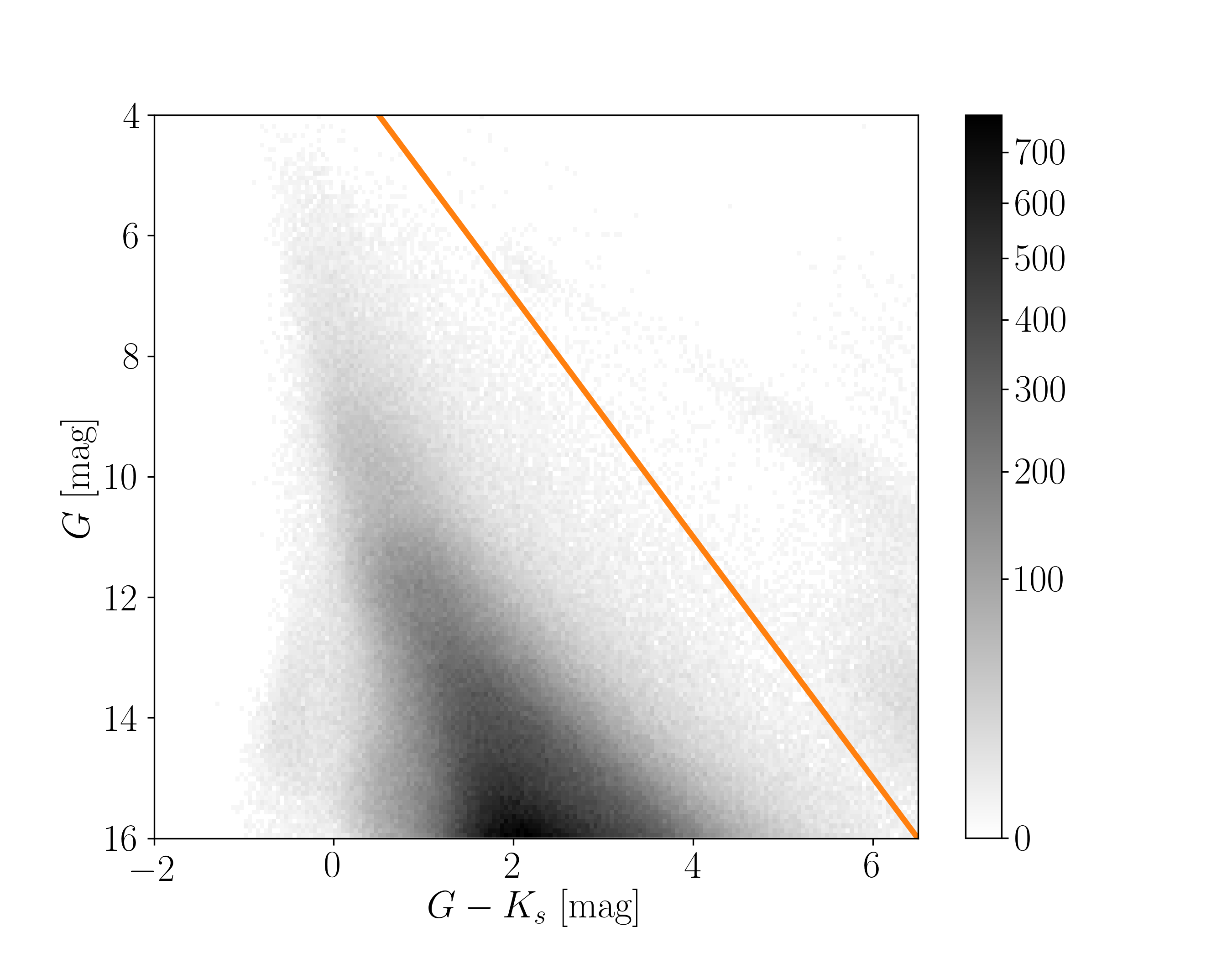}
    \caption{$G - K_s$ vs. $G$ colour-magnitude diagram of the sources selected after applying the criteria in the $(J-K_s)_0$ vs $G-K_s$ and $G-K_s$ vs $J-H$ colour-magnitude diagrams (see text). The solid orange line has equation $G = 2(G - K_s) + 3$, and aims at excluding the few  cool giants that may still contaminate the sample.}
    \label{fig:cmd3}
\end{figure}

Fig. \ref{fig:cmd2} shows the distribution of sources in the $G-K_s$ vs. $J - H$ colour-colour diagram.
O and B-type stars lie on a sequence in the $J - H$ vs. $G - K_s$  colour-colour diagram as a consequence of interstellar reddening and are clearly separated from redder turn-off stars and giants. We therefore select stars located in the region defined by the  equations: 
\begin{align}\label{Eq:j-h}
&J - H < 0.15(G-K_s) + 0.05 \nonumber \\
&\mathrm{and} \nonumber \\
&J - H > 0.15(G-K_s) - 0.15    
\end{align}
(solid grey lines in Fig. \ref{fig:cmd2}). 
Finally, we select stars in the  $G$ vs. $G - K_s$ colour-magnitude diagram, where giants still contaminating our sample can be easily separated from OBA stars (see Fig. \ref{fig:cmd3}). In particular we select stars satisfying:
\begin{equation}
G > 2(G - K_s) + 3     
\end{equation}
(solid grey line, in Fig. \ref{fig:cmd3}). 
The catalogue consists of 988 202 entries.

\subsubsection{Characteristics of the target sample}\label{sec:characteristics}
\begin{figure*}
    \centering
    \includegraphics[width = \hsize]{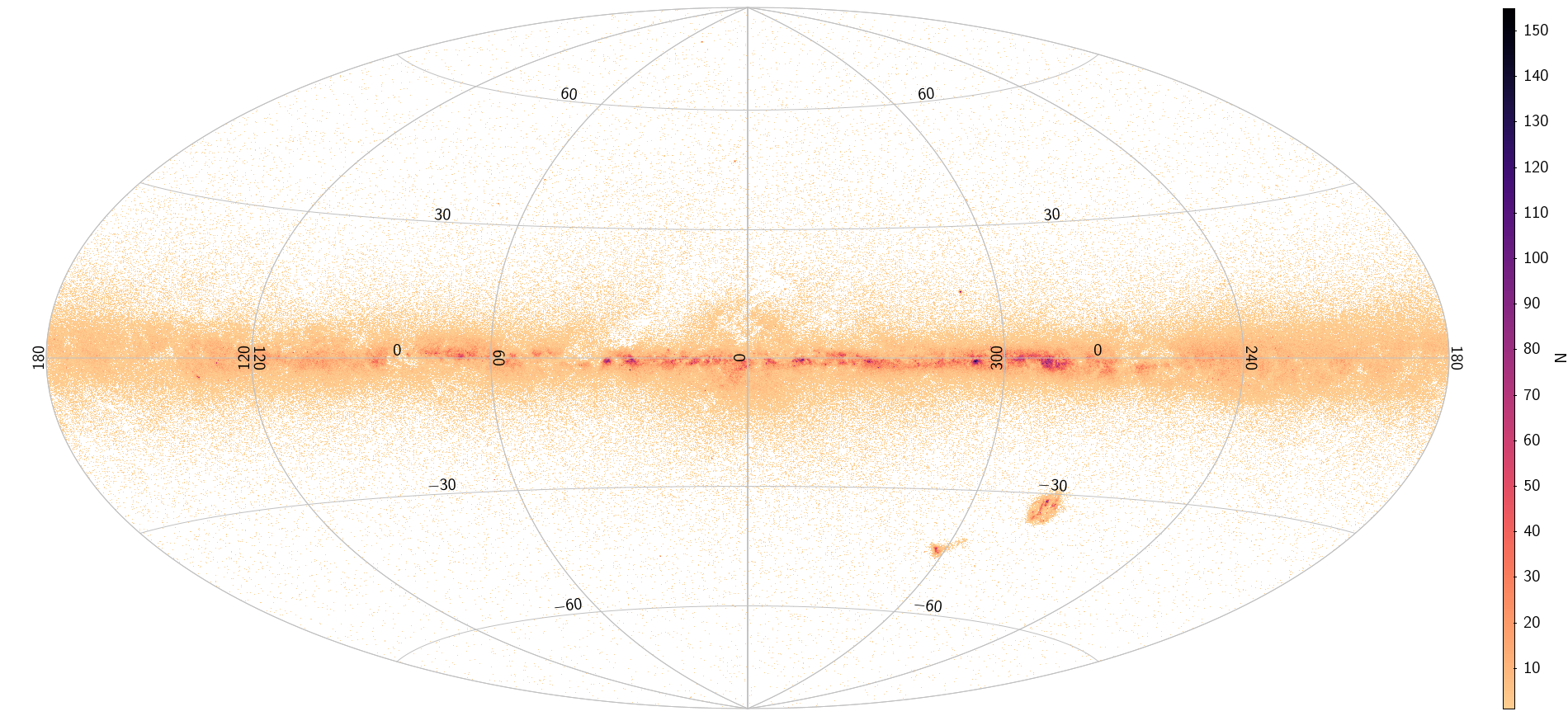}
    \caption{On-sky density distribution of the candidate OBA stars selected in Section \ref{sec:data}, in Galactic coordinates. Most of the sources are located in the Galactic plane ($b = 0^{\circ}$) and in the Large and Small Magellanic clouds. }
    \label{fig:sky_plot}
\end{figure*}

Fig. \ref{fig:sky_plot} shows the sky distribution of the sample selected in the previous Sections in Galactic coordinates $(l, b)$.
The stars in our catalogue are distributed in our Galaxy (mainly on the Galactic plane) and in the Large (LMC, $l,b \sim 280^{\circ},-34^{\circ}$) and the Small  (SMC, $l, b \sim 303^{\circ}, -45^{\circ}$) Magellanic Clouds. We stress that massive stars in the Magellanic clouds are included in the target sample of the SDSS-V survey. In this paper however we focus on the Milky Way disc, thus in the following sections we restrict our sample to stars with $|b| < 25^{\circ}$.
Massive star forming regions can be recognised as over-densities in the source distribution. 

Dust features are visible as 'gaps' in the star distribution. For example, the Aquila rift, the Pipe Nebula and the Ophiucus clouds can be easily identified towards the Galactic centre ($l, b = 0, 0$ deg). 
By visually inspecting the features of Fig. \ref{fig:sky_plot}, it is possible to identify old stars  contaminating the sample. For instance,  at $l, b \sim 309^{\circ},15^{\circ}$  the $\omega$ Centaurus globular cluster is visible, while the stars towards the Galactic centre delineate the shape of the bulge. We further study the purity and completeness of our sample in Sec. \ref{sec:completeness}.

\begin{figure*}
    \centering
    \includegraphics[width = 0.45\hsize]{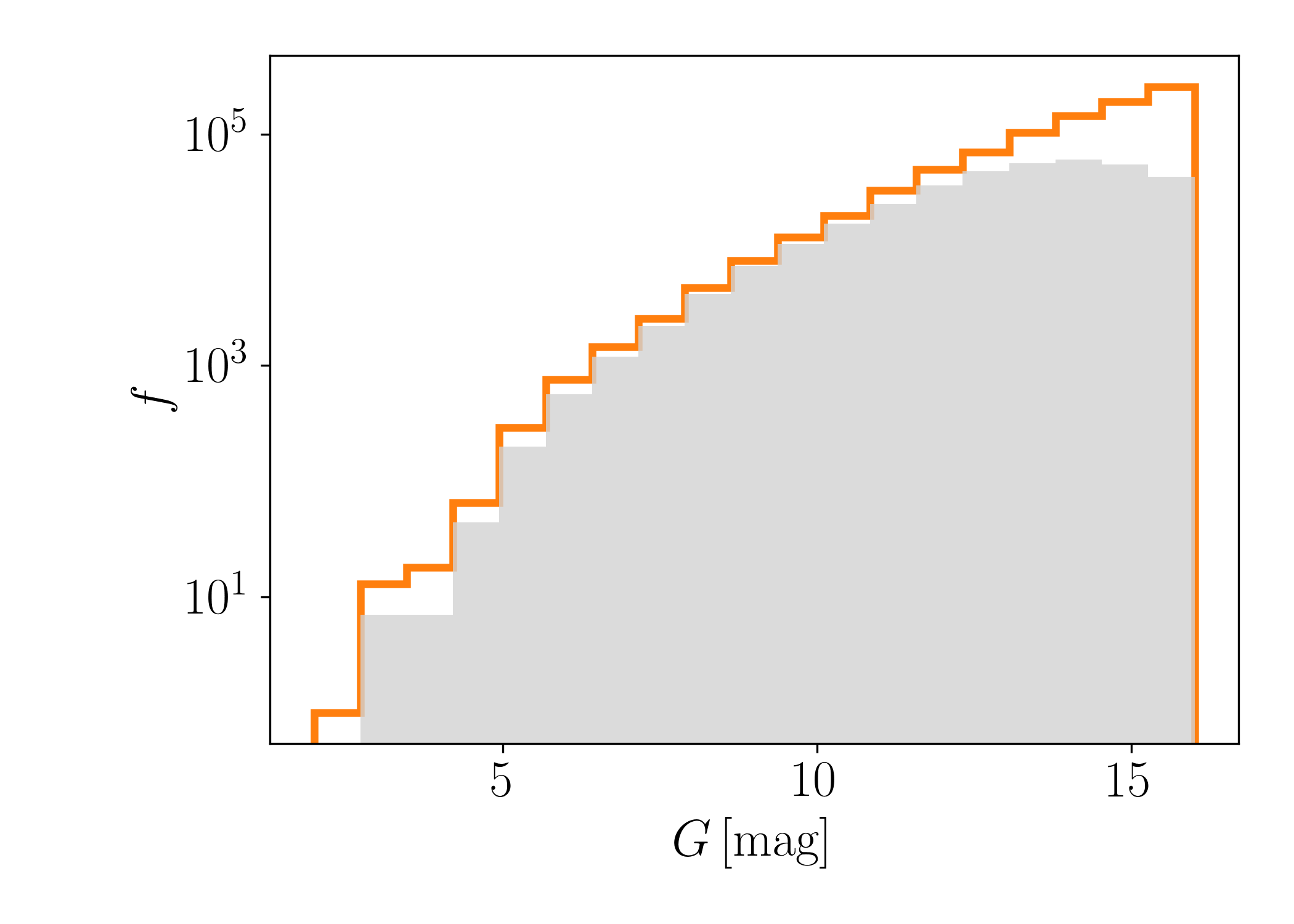}
    \qquad
    \includegraphics[width =0.45\hsize]{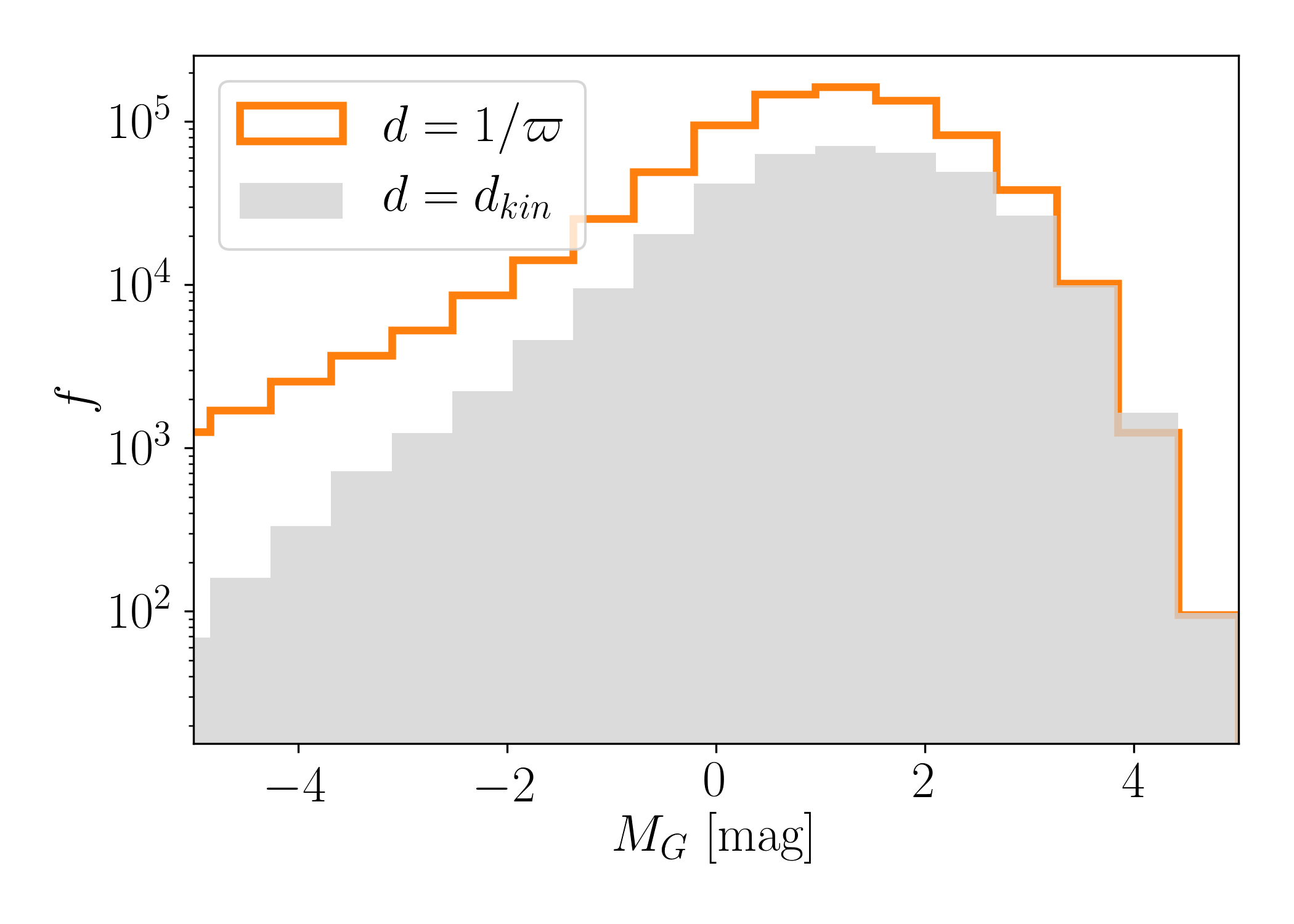}
    \caption{Left: $G$ magnitude distribution of all the OBA stars selected in Section \ref{sec:selection} (orange histogram) and of the sources selected in Section 3, with astro-kinematic distances comparable with those estimated by \cite{Bailer-Jones2018} (grey histogram). Right: absolute magnitude distribution of all the OBA stars selected in Section \ref{sec:selection}, computed by using their parallax (orange histogram), and of the sources selected in Section 3, computed by using their astro-kinematic distances (grey histogram).}
    \label{fig:mag_hist}
\end{figure*}

Fig. \ref{fig:mag_hist} shows the distribution in apparent (left) and absolute magnitude (right) of the sources. The orange histograms show the distribution of all the sources in our sample. The grey histograms show the distribution of the sources selected in Section \ref{sec:dist_estimates}. On the right hand panel, absolute magnitudes are computed by using the inversion of parallaxes as the distance (orange histogram) or astro-kinematic distances (grey histogram, see Section \ref{sec:dist_estimates}). The grey distribution appears shifted towards larger $M_G$ values. This is due to the fact that, for stars with small parallaxes (and large parallax errors), our astro-kinematic distances are shorter than those obtained by inverting the parallax. 
\begin{figure*}
    \centering
    \includegraphics[width = 0.45\hsize]{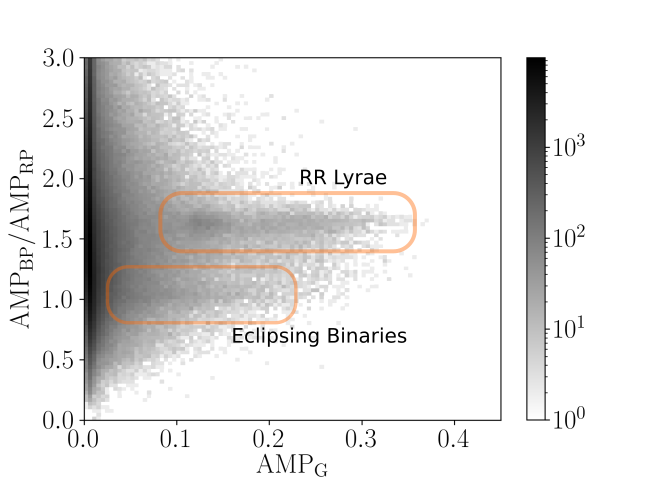}
    \includegraphics[width =0.45\hsize]{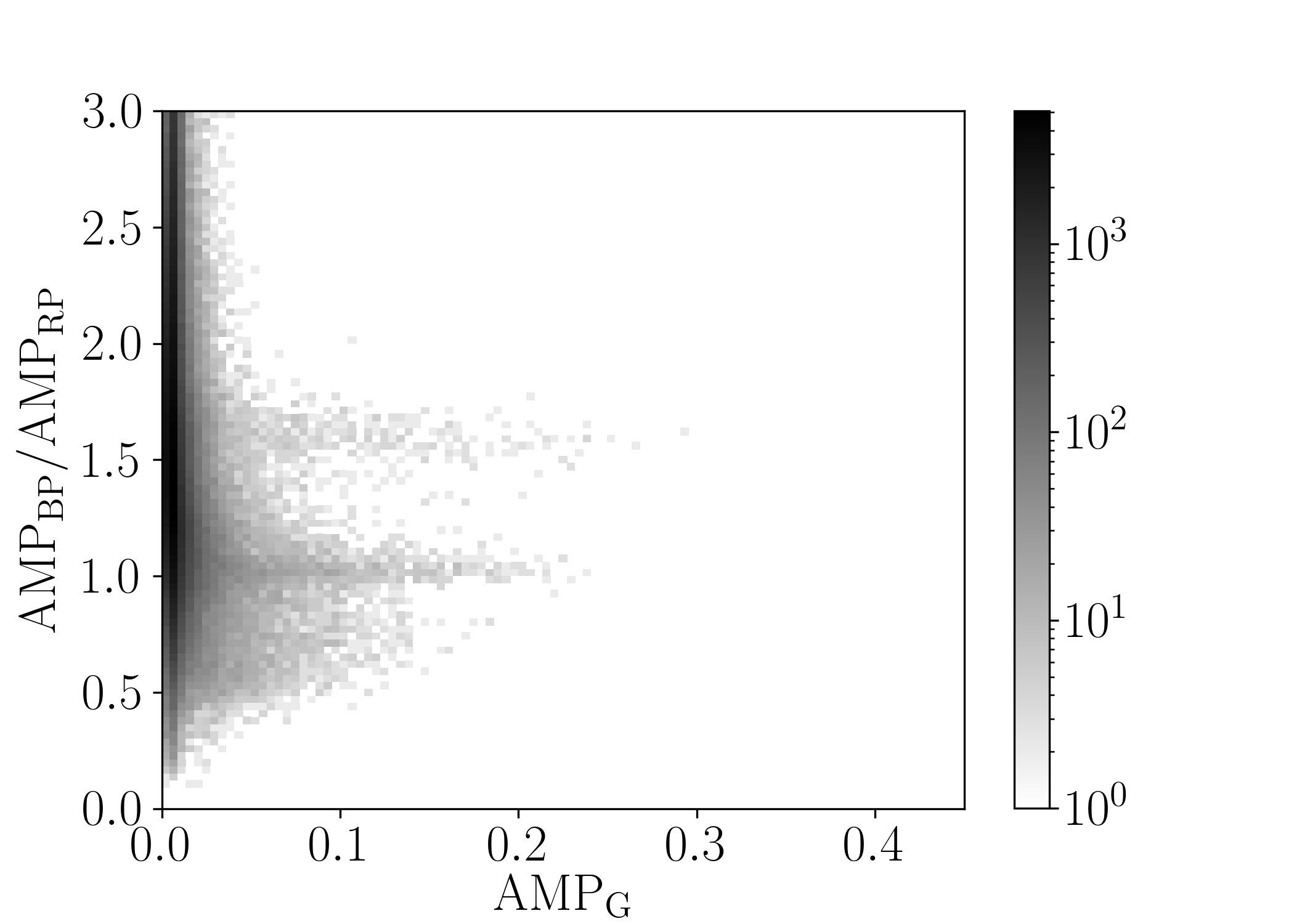}
    \caption{$G-$band variability vs. $G_{\mathrm{RP}}/G_{\mathrm{BP}}$ variability of all the sources selected in Section \ref{sec:selection} (left) and of the sources selected in Section \ref{sec:filtered_sample} (right). Eclipsing binaries have the same level variability in $G_{\mathrm{BP}}$ and $G_{\mathrm{RP}}$  and $G-$variability lower than 0.3. RR Lyrae stars have $G_{\mathrm{BP}}/G_{\mathrm{RP}}$ variability $\sim 1.6$  and  $G-$band variability lower than 0.4.}
    \label{fig:variability}
\end{figure*}

Fig. \ref{fig:variability} (left) shows the variability distribution of the sources in the $G$ band vs. the ratio of the variability in the $G_{\mathrm{BP}}$ and $G_{\mathrm{RP}}$ bands. Following roughly \cite{Belokurov2017}, \cite{Deason2017} and \cite{Iorio2018}, the variability in a given filter $x$ is defined as:
\begin{equation}
   \mathrm{AMP_x}  = \frac{\sqrt{\texttt{phot\_x\_n\_obs}}}{\texttt{phot\_x\_mean\_flux\_over\_error}},
\end{equation}
where  \texttt{phot\_x\_n\_obs} is the number of observations contributing to the $x$-band
photometry, and 
\begin{align}
&\texttt{phot\_x\_mean\_flux\_over\_error} = \nonumber\\ &\texttt{phot\_x\_mean\_flux}/\texttt{phot\_x\_mean\_flux\_error}.
\end{align}
While most of our sources are not variable, some cluster in two regions:
RR Lyrae stars have $G$-band variability ranging from 5 to 40\% \citep{Belokurov2017} and $G_{\mathrm{BP}}/G_{\mathrm{RP}}$ variability around 1.6; eclipsing binaries have $G$ variability up to 30\% and roughly equal $G_{\mathrm{BP}}$ and $G_{\mathrm{RP}}$ variability. Fig. \ref{fig:variability} (right) shows the same as Fig. \ref{fig:variability} (left) but for the sources selected in Section 3 (see below), that have astro-kinematic distances compatible with those estimated by \cite{Bailer-Jones2018}. This selection removes RR Lyrae stars, while retaining eclipsing binaries: this is expected as RR Lyrae stars do not follow the same kinematics as the OBA stars that we are interested in.

\subsubsection{Completeness and purity}\label{sec:completeness}
\begin{figure*}
    \centering
    \includegraphics[width = 0.49\hsize]{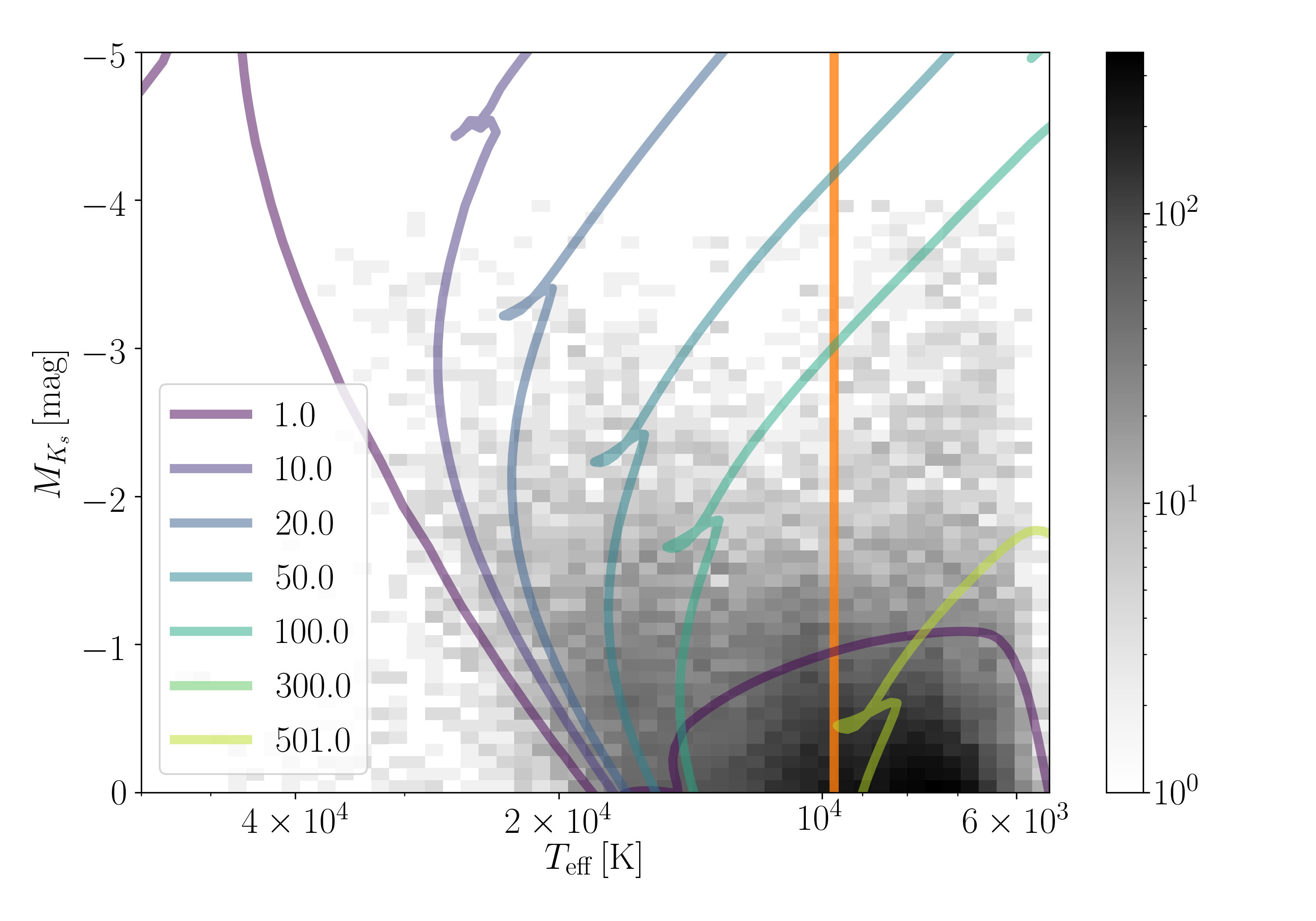}
    \includegraphics[width = 0.49\hsize]{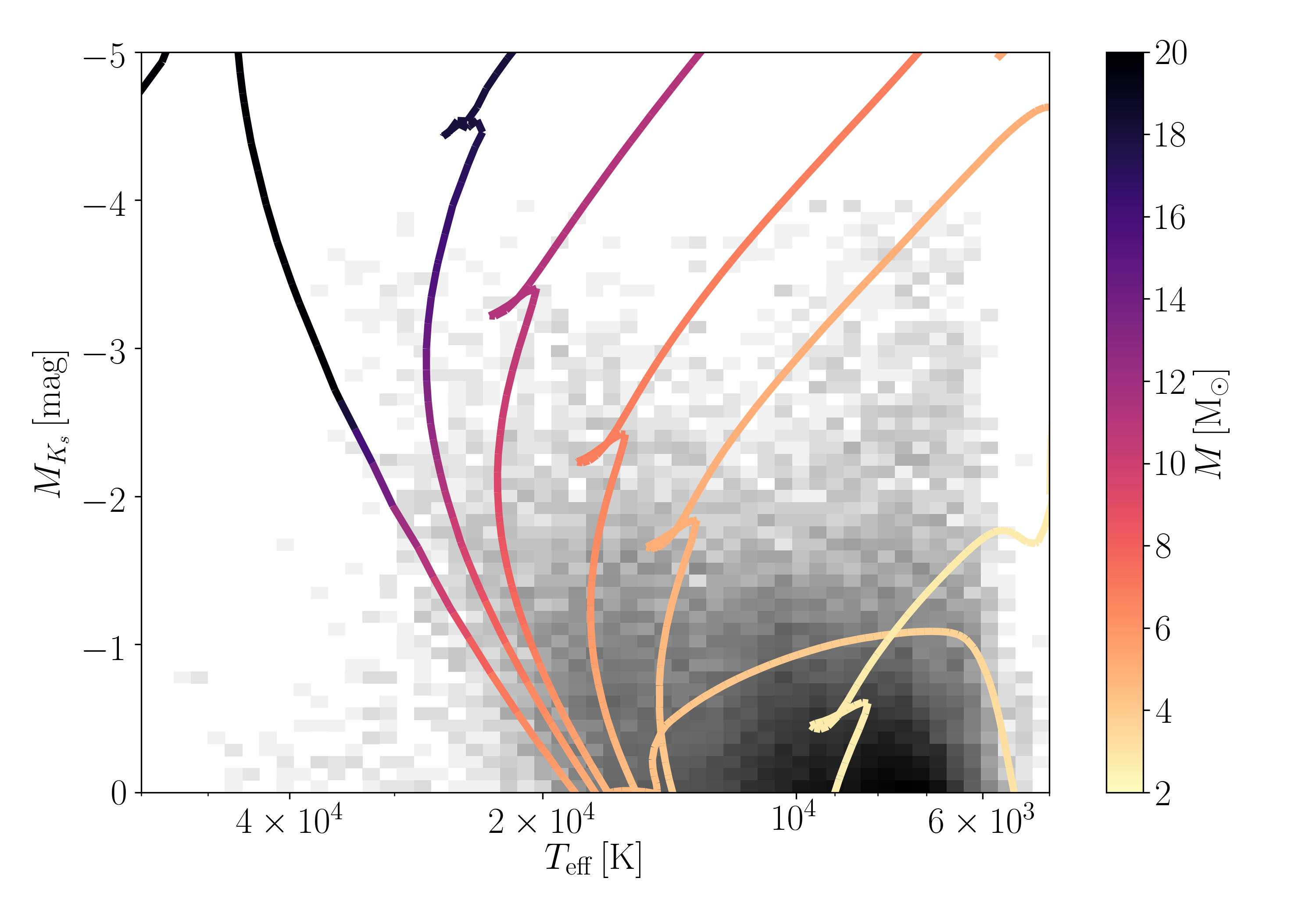}
    \caption{Left: Distribution of the sources with LAMOST DR6 spectra in the $\log T_{\mathrm{eff}}$ vs. $M_{K_s}$ (gray). The orange solid line indicates $T_{\mathrm{eff}} = 9700$ K, which corresponds to the temperature of a A0V-type star \citep{Pecaut2013}.  The isochrones are colour-coded according to their age (in Myr). Right: same as left, with the isochrones colour-coded according to their mass.}
    \label{fig:lamost}
\end{figure*}

To study the purity level of our sample we cross-matched it with LAMOST DR6 (\url{http://dr6.lamost.org/}), and find 36 617 sources in common.
We derived effective temperatures ($T_{\mathrm{eff}}$) and surface gravity ($\log g$) from the LAMOST spectra by fitting ab initio model spectra generated with the 1D-LTE ATLAS12 model atmosphere (Xiang et al., in prep).
Fig. \ref{fig:lamost} shows the distribution of sources in the $\log T_{\mathrm{eff}}$ vs. $M_{K_s}$ plane. Around the 45\% of the sources are hotter than 9 700 K (orange solid line in Fig. \ref{fig:lamost}, left, corresponding to the temperature of a A0V-type star) and around the 20\%  are hotter than 14 000 K (roughly the temperature of a B7V-type star).  We assume that we can extrapolate the  same numbers to the entire sample.  
The PARSEC isochrones ($A_V = 0$ mag and solar metallicity) in Fig. \ref{fig:lamost} \citep{Bressan2012, Tang2014} show that our main source of contamination are evolved massive or intermediate mass stars (e.g. yellow and blue super-giants) and that the sample is substantially free from RGB and AGB stars, which are colder and more luminous than the stars in our sample. In the left-hand panel, the isochrones are colour-coded by their log age (from 1 Myr to 500 Myr), in the right hand panel they are colour-coded by mass. 

To estimate the completeness of the sample we cross-matched with the O and B-type star catalogue by \cite{Liu2019} and the O-type star catalogue by \cite{Sota2014} and \cite{Maiz2016}. \cite{Liu2019}  sample consists of 16 032 stars that reduce to 15 206 after cross-matching with \textit{Gaia} EDR3, 2MASS and further checking for duplicated sources (1" cross-match radius), and to 9 083 stars by applying Eq. \ref{eq:MK0}. The cross-match between ours and \cite{Liu2019} catalogue gives 8212 sources. The stars that are not included in the cross-match are those whose infrared colours are not consistent with our selection.
\cite{Sota2014} and \cite{Maiz2016} sample (the Galactic O-star spectroscopic survey, GOSSS) consists of 590 stars, that reduce to 580 after cross-matching with \textit{Gaia} EDR3 and 2MASS. Our catalogue contains 503 ($\sim$86\%) of the GOSSS stars. As with the LAMOST sample, the stars missing from our selection do not follow some of our photometric selection criteria. For example many of the missing GOSSS sources (49\%, 54 stars) have $J - H >  0.15 (G - K_s) + 0.05$ (see Eq. \ref{Eq:j-h}), that is 2MASS-\textit{Gaia} EDR3 colours consistent with being giants.

\section{Distance estimates}\label{sec:dist_estimates}
\begin{figure*}
    \centering
    \includegraphics[width = 0.48\hsize]{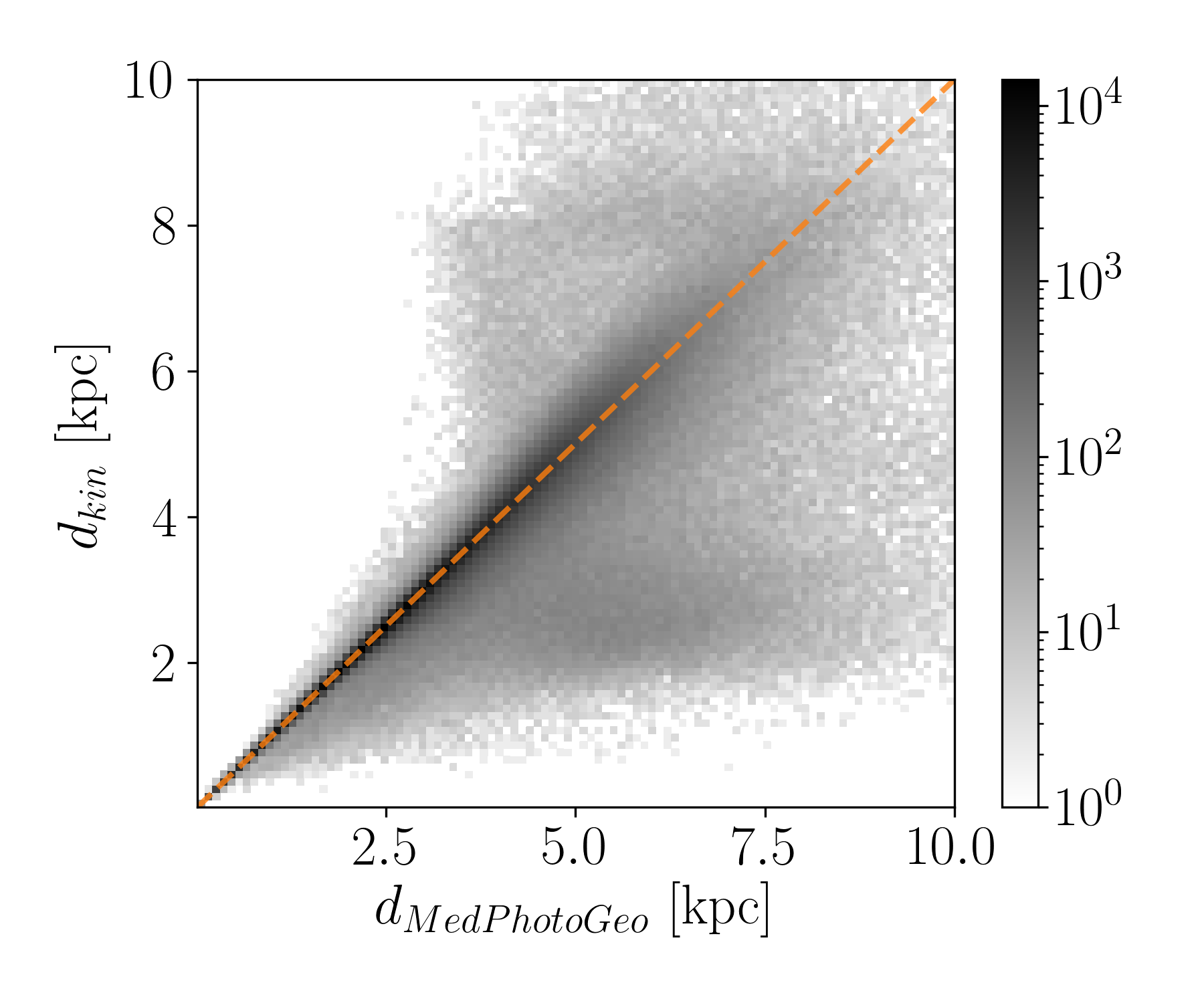}
    \includegraphics[width = 0.48\hsize]{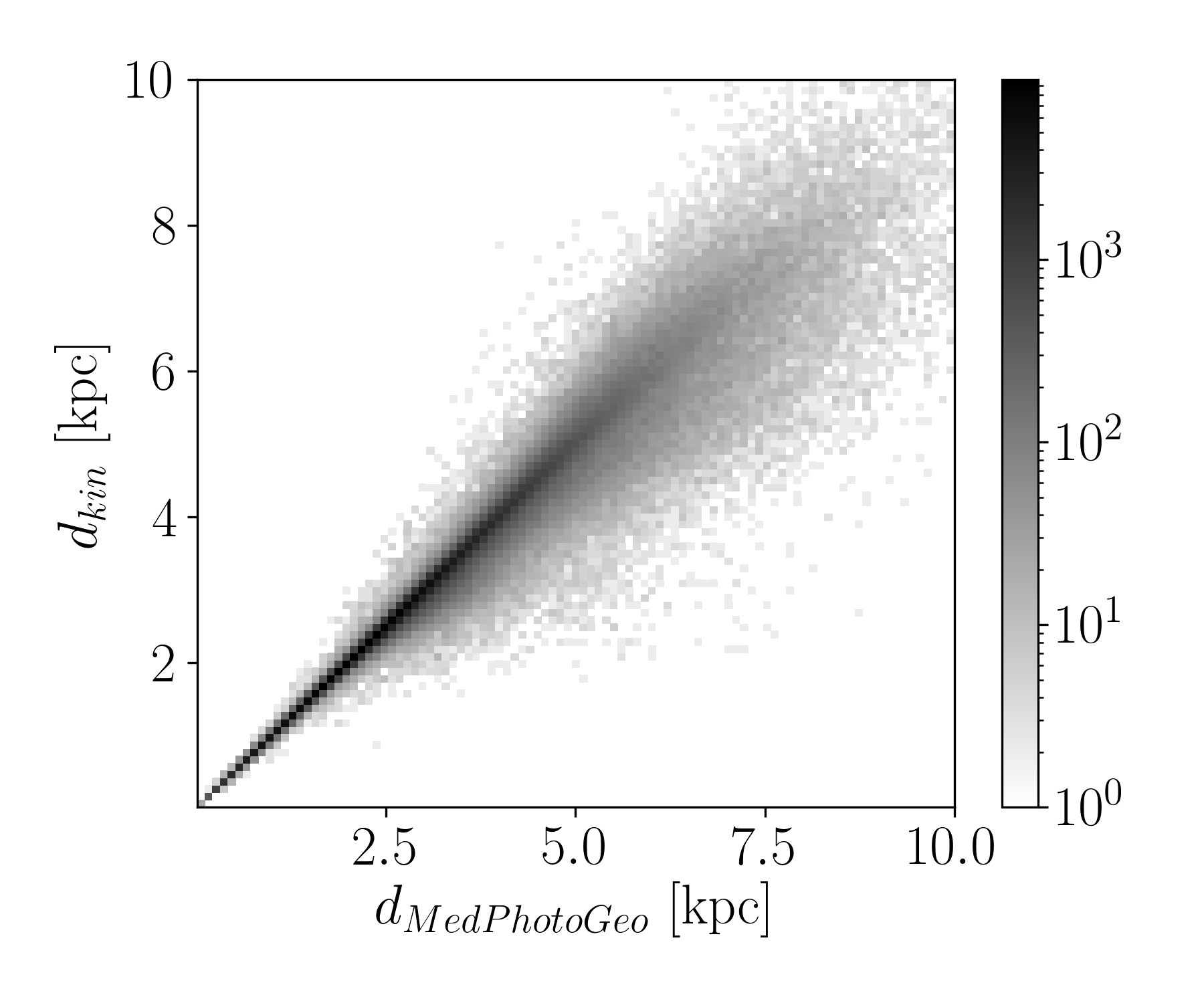}
    \caption{Comparison between the photo-geometric distances estimated by \cite{Bailer-Jones2020} and our astro-kinematic distances (see Sec. \ref{sec:dist_estimates}), after removing spurious sources (left), and after removing sources with vertical velocity not consistent with our kinematic model and with $M_{K_s} < 0 \, \mathrm{mag}$ (right). 
    The orange dashed line has equation $y = x$.}
    \label{fig:dkin_dBJ}
\end{figure*}

To study the 3D space distribution of our sources, precise distance estimations are needed. 
We estimate distances by using a  model designed to reproduce the properties of our data-set in terms of spatial and luminosity distribution, and the additional information that stars belonging to our sample should follow Galactic rotation, with a small, typical velocity dispersion. By making such assumption we neglect non-circular streaming motions. 
We can then predict the \textit{true} proper motions ($\mu'_l, \mu'_b$) of the stars in our sample and compare them with the observed ones, thus adding a further constraint to the distance estimation and deriving 'astro-kinematic' distances. The probability density function (\textit{pdf}) for a star in our sample to be at a certain distance $d_{kin}$ is given by:
\begin{align}\label{eq:pdf_text}
   & p(d~|~ \Vec{o},  m_{K_s}, \Theta_{\mathrm{KM}}, \Theta_{\mathrm{SM}}, \Theta_{\mathrm{CMD}})  \propto  \nonumber\\
  & p(\Vec{o} ~|~ \Theta_{KM}) p(d, m_{K_s} ~|~ l, b, \Theta_{\mathrm{SM}}, \Theta_{\mathrm{CMD}}),
\end{align}
where 
\begin{itemize}
\item $\Vec{o } =  \left(\varpi - \varpi_0, \mu_{l*},  \mu_b \right)$ 
is the array of the astrometric observables, i.e. parallax $\varpi$ (with $\varpi_0$ the parallax zero-point) and proper motions components in $l$ and $b$, $\mu_l*$ and $\mu_b$;
\item $m_{K_s}$ is the apparent magnitude in the $K_s$ band;
\item $\Theta_{KM}$ represents our kinematic model;
\item $\Theta_{SM}$ represents our model for the distribution of stars in the Galaxy;
\item $\Theta_{CMD}$ accounts for the observational effects due to our selection function on the spatial distribution of our sample. 
\end{itemize}
The method and all the terms in Eq. \ref{eq:pdf_text} are described in detail in Appendix \ref{sec:kin_dist}. We adopt as the astro-kinematic distance estimate $d_{kin}$ the mode of the \textit{pdf} of Eq. \ref{eq:pdf_text}, and we use the 16\textit{th} and 84\textit{th} percentiles to estimate distance errors. We also tried using the median of the \textit{pdf} as a distance estimate, but this did not cause significant differences in the maps presented in Section \ref{sec:results}.
Figure \ref{fig:dkin_dBJ} (left) shows the comparison between our astro-kinematic distances and the photo-geometric distances estimated by \cite{Bailer-Jones2020}. More than the 85\% of the sources have astro-kinematic distances consistent  within 1$\sigma$ with the photo-geometric distances from  \cite{Bailer-Jones2020}. We comment on the sources with inconsistent distances in Section \ref{sec:kinematic cleaning}.

\section{Filtered sample}\label{sec:filtered_sample}
In this Section we define a sub-set of the target sample (Section \ref{sec:data}), which we use to map the structure of the young Milky Way disc. Such "filtered" sample is obtained by cleaning the target sample for sources with likely spurious astrometric solutions (Sec. \ref{sec:spurious_sources}) and by removing sources with kinematic properties not consistent with the model that we used to estimate astro-kinematic distances (Sec. \ref{sec:kinematic cleaning}).

\subsection{Removal of spurious sources}\label{sec:spurious_sources}
\begin{figure}
    \centering
    \includegraphics[width = \hsize]{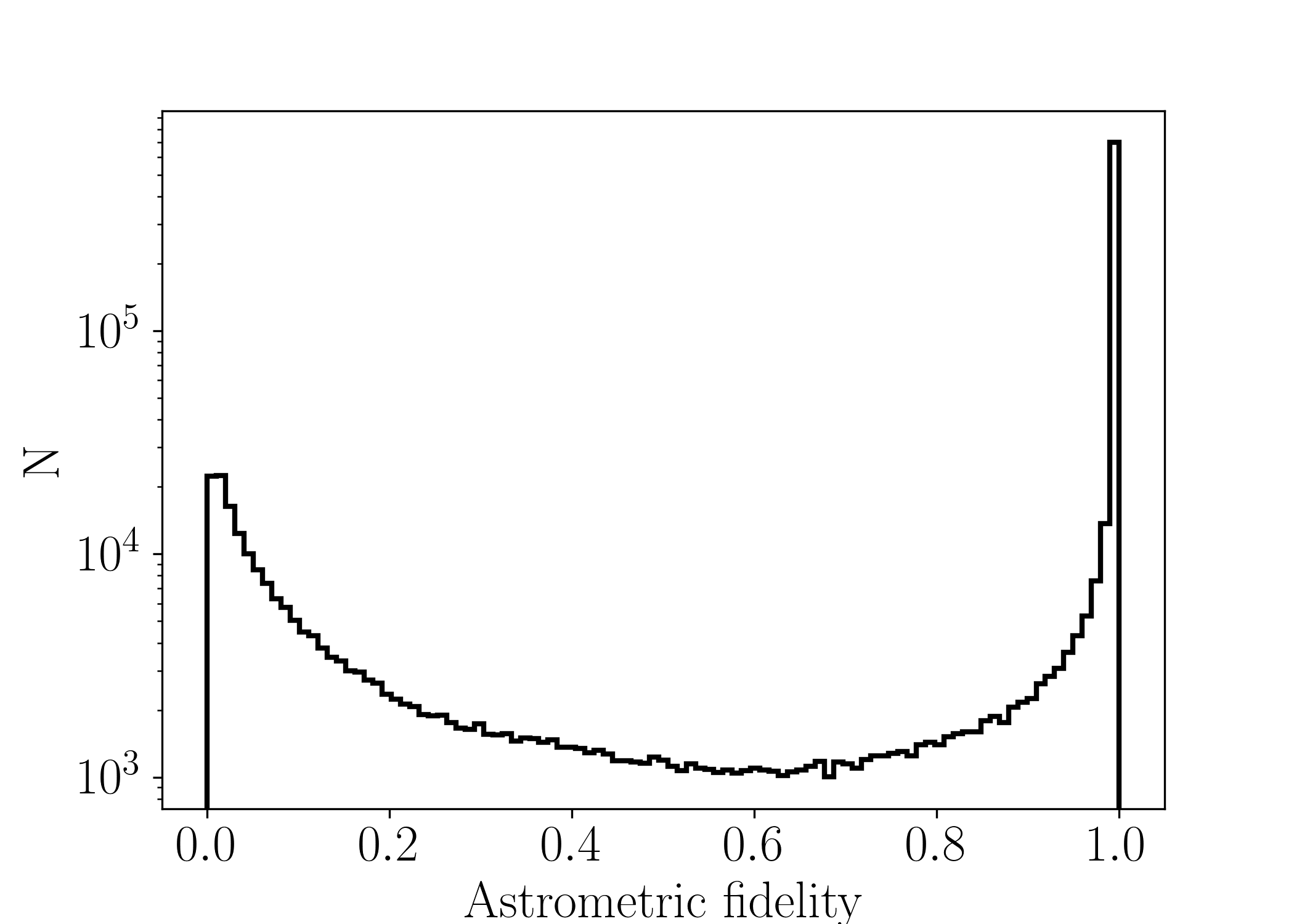}
    \caption{Astrometric fidelity distribution for the sources in our target sample. 
    Low astrometric fidelity values ($< 0.5$) correspond to bad astrometric solutions, high astrometric fidelity values ($>0.5$) correspond to good astrometric solutions. }
    \label{fig:probability}
\end{figure}

To remove spurious sources, we use a method developed by  \cite{Rybizki2021}, based on \cite{Smart2020}. Spurious sources have poor astrometric solutions that can be due to the inconsistent matching of the observations to different physical sources. This is more likely to occur in regions of high source surface density (for example, in the Galactic plane) or for close binary systems (either real or due to perspective effects).
To identify poor astrometric solutions in their solar neighbourhood catalogue, \cite{Smart2020} constructed a random-forest classifier that assigns a probability to each source of having a good (or bad) astrometric solution, based on astrometric quantities and quality indicators. The classification probability is $\sim$0 for sources with poor astrometric solutions, and $\sim$1 for sources with good astrometric solutions. \cite{Smart2020} applied this classifier to stars with distances less than 100~pc. \cite{Rybizki2021} used similar methods to develop a neural network classifier, which they applied to the entire \textit{Gaia} EDR3 catalogue. 
The distribution (in logarithmic scale) of "astrometric fidelities" for the sources in our target sample is shown in Fig. \ref{fig:probability}. We select sources with astrometric fidelities $>$ 0.5 (around 75\% of the target sample), following the definition of "good" and "bad" astrometric solutions in \cite{Rybizki2021}. The distribution in the Galactic plane of sources with good and bad astrometric solutions is shown in Fig. 8 of \cite{Rybizki2021}.
While this procedure allows to obtain a more reliable sample, it might also eliminate binaries.
Binaries however do not constitute the the focus of this paper and do not affect the space distribution of our filtered sample.

\subsection{Kinematic cleaning}\label{sec:kinematic cleaning}
Our astro-kinematic distance estimates 
are not accurate for sources that do not follow the disc kinematics assumed by our model. 
Such sources can be grouped in two categories:
\begin{enumerate}
\item stars whose kinematics is influenced by the bar. They are located mostly towards the Galactic Centre, and have astro-kinematic distances larger than Bailer-Jones photo-geometric distances. These stars could in principle be young. Their astro-kinematic distance estimates could however be wrong due to the kinematic effects of the bar itself, that have not been included in our model (see Section \ref{sec:dist_estimates} and Appendix \ref{sec:kin_dist}). In the following Sections we focus on the in-plane distribution of stars within $d_{kin} \lesssim$ 5 kpc, and thus we do not consider sources in the bar/bulge regions.

\item stars with heated vertical kinematics.
These stars have Bailer-Jones photo-geometric distances $d_{BJ} \gtrsim 3 \, \mathrm{kpc}$ and astro-kinematic distances $d_{kin} \sim 2 \, \mathrm{kpc}$.
They could correspond to old contaminants, such as the RR Lyrae-type variable stars discussed in Section \ref{sec:characteristics}.
To remove them we proceed as follow.  First, we compute their vertical velocity, $v_z$ (see  Appendix \ref{sec:vertical velocity}).We fit the vertical velocity distribution with a Gaussian Mixture model with two components, the first (component A) with mean vertical velocity $<v_z>_A = -0.5 \, \mathrm{km \, s^{-1}}$   and velocity dispersion $\sigma_{v_z, A} = 7.0 \, \mathrm{km \, s^{-1}}$ which has properties comparable with the young population that we are interested in studying, and the second (component B), with $<v_z>_B = 1.6 \, \mathrm{km \, s^{-1}}$  and velocity dispersion $\sigma_{v_z,B} = 29. \, \mathrm{km \, s^{-1}}$,  which instead better reproduces the properties of an old population. We estimate the probability for each star to belong to either component A or B, and we select those stars with larger probability of belonging to component A.
\end{enumerate}

Figure \ref{fig:dkin_dBJ} (right) shows the comparison between the astro-kinematic distances and Bailer-Jones photo-geometric distances after cleaning the target sample as explained above. At this point, around the 95\% of the sources have astro-kinematic distances consistent with Bailer-Jones photo-geometric distances within 1$\sigma$.

Finally, by using our astro-kinematic distances, we estimate the absolute magnitude of each star in the $K_s$ band, $M_{K_s}$, and we select stars with $M_{K_s} < 0$ mag. This condition refines Eq. \ref{eq:MK0}, and aims at selecting stars of type earlier than B7V.

After applying the quality cuts described in this Section, our filtered sample reduces to 435 273 stars.

\section{3D space distribution}\label{sec:results}
    To create the 3D density maps, we follow the method outlined in \cite{Zari2018}. We compute galactic Cartesian coordinates, $x$, $y$, and $z$, for all the sources by using our newly estimated astro-kinematic distances and we define a box $V=12 \times12\times1.5 \, \mathrm{kpc}$ centred on the Sun. We divide the cube into volume elements $v = 5 \times 5 \times 5 \, \mathrm{pc}$.  After computing the number of stars in each volume, we estimate the star density $D(x,y,z)$ by smoothing the distribution by means of a 3D Gaussian filter, using a technique similar to that used by \cite{Bouy2015}.The Gaussian width (equal on the three axes) is $w=5$ pc the Gaussian is truncated at $3\sigma$. The choice of a certain $w$ value is arbitrary. A high $w$ value produces a smooth, less detailed map, while a low $w$ value results in a noisy map.

\begin{figure*}
    \centering
    \includegraphics[width = \hsize]{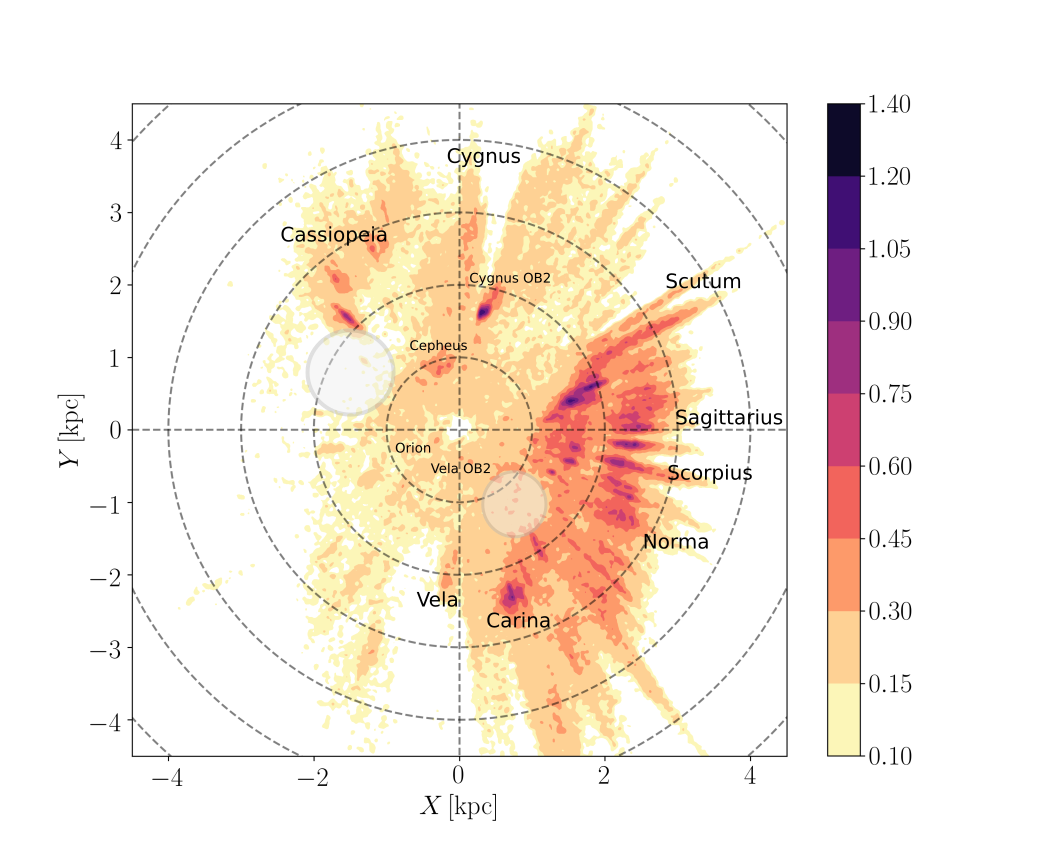}
    \caption{Surface density of the stars selected in Section 3 projected on the Galactic Plane.
    The Sun is in (0,0), the x-axis is directed towards the Galactic centre, and the y-axis towards Galactic rotation.
    The z-axis is perpendicular to the plane. 
    The density is displayed in arbitrary units. 
    The dashed circles have radii from 1 to 6 kpc, in steps of 1 kpc. 
    The labels follow  the nomenclature proposed by K. Jardine (http://gruze.org/galaxymap/map\_2020/). 
    The gaps in the density distribution mentioned in the Section \ref{sec:results} are indicated by the grey circles.} 
    \label{fig:map0}
\end{figure*}

Fig. \ref{fig:map0} shows the projection on the Galactic plane of the filtered sample selected in Section \ref{sec:filtered_sample}. A number of features is visible, showing different structures at different scales and distances.
At the centre of the map ($x, y \sim 0, 0$), a low-density gap is visible. This is due to a combination of reasons. First, in the solar neighbourhood the number of OBA stars is low. The Sun is indeed currently located in a cavity, partially filled with hot, low-density gas, which is usually called the Local Bubble (LB). The LB was likely created 10-20 Myr ago by a number of supernova explosions that likely halted any low- or high-mass star formation episode in the region. Second, our selection criteria exclude by construction stars of later spectral type than $\sim$B7V. Thus, even if some late type B stars present within $200-300 \, \mathrm{pc}$ from the Sun, they are excluded from our selection. Third, selection effects due to the fact that a) nearby bright sources are not included in \textit{Gaia} EDR3 and b) some of those that are included might have poor 2MASS photometry, and thus are excluded by the colour selection in Eq. \ref{eq:3}.

The small dense clumps can be associated to well studied OB associations, for example Cygnus, Carina, Cassiopeia, and Vela, are visible. The distribution of the lower density contours traces the Milky Way young disc structure, in particular the spiral arm location (see Sec. \ref{sec:ob} and \ref{sec:discussion}). The density distribution presents numerous low-density gaps.  These could be due to the fact that our view might be obscured by interstellar dust towards certain lines of sight, however some gaps are located in regions of relatively low extinction (see also Section \ref{sec:dust_maps}). For example the Perseus gap ($x,y \sim -2, 1 \, \mathrm{kpc}$) has been detected in the  distribution of  other young  tracers, which are traditionally associated with spiral arms, such as HII maps (Hou \& Han, 2014), Cepheid distribution \citep{Skowron2019}, see Sec. \ref{sec:cepheids}) as well as in the OB stars shown by \cite{Romero2019} and \cite{Poggio2018} (see Sec. \ref{sec:ob}), and the high-mass star-forming regions of \cite{Reid2014} and \cite{Reid2019} (see Sec. \ref{sec:reid}). By tracing the open cluster distribution in the Galactic disc, \cite{Cantat2020} suggests that a similar gap might be observed around $x,y \sim 1, -1 \, \mathrm{kpc}$, where our map show an under-density, although less pronounced than the Perseus gap. In general, the distribution of stars along the "arms" is not homogeneous, but characterised by under- and over-densities. For distances larger than 3-4 kpc the number of sources (and thus their spatial density) decreases due to our magnitude limit at $G = 16 \, \mathrm{mag}$ (see Sec. 2).

The radial features are  caused by 'shadow cones' produced by foreground extinction, and by distance uncertainties. As shown in Fig. \ref{fig:frac_uncertainty}, distance uncertainties are lower than 10\% for stars within 5~kpc, and therefore are only partially responsible for the elongation.

\subsection{The Galactic warp}
As mentioned in the Introduction, early-type star samples have been used to study the properties of the Galactic warp \citep[see][and references therein]{Poggio2018, Romero2019}. Although in this paper we do not focus specifically on the warp, we studied the median height of our filtered sample with respect to the plane of the Galaxy and the median vertical velocity distribution, and compared our results to those by \cite{Poggio2018} and \cite{Romero2019} to further validate our selection.

\begin{figure}
    \centering
    \includegraphics[width = \hsize]{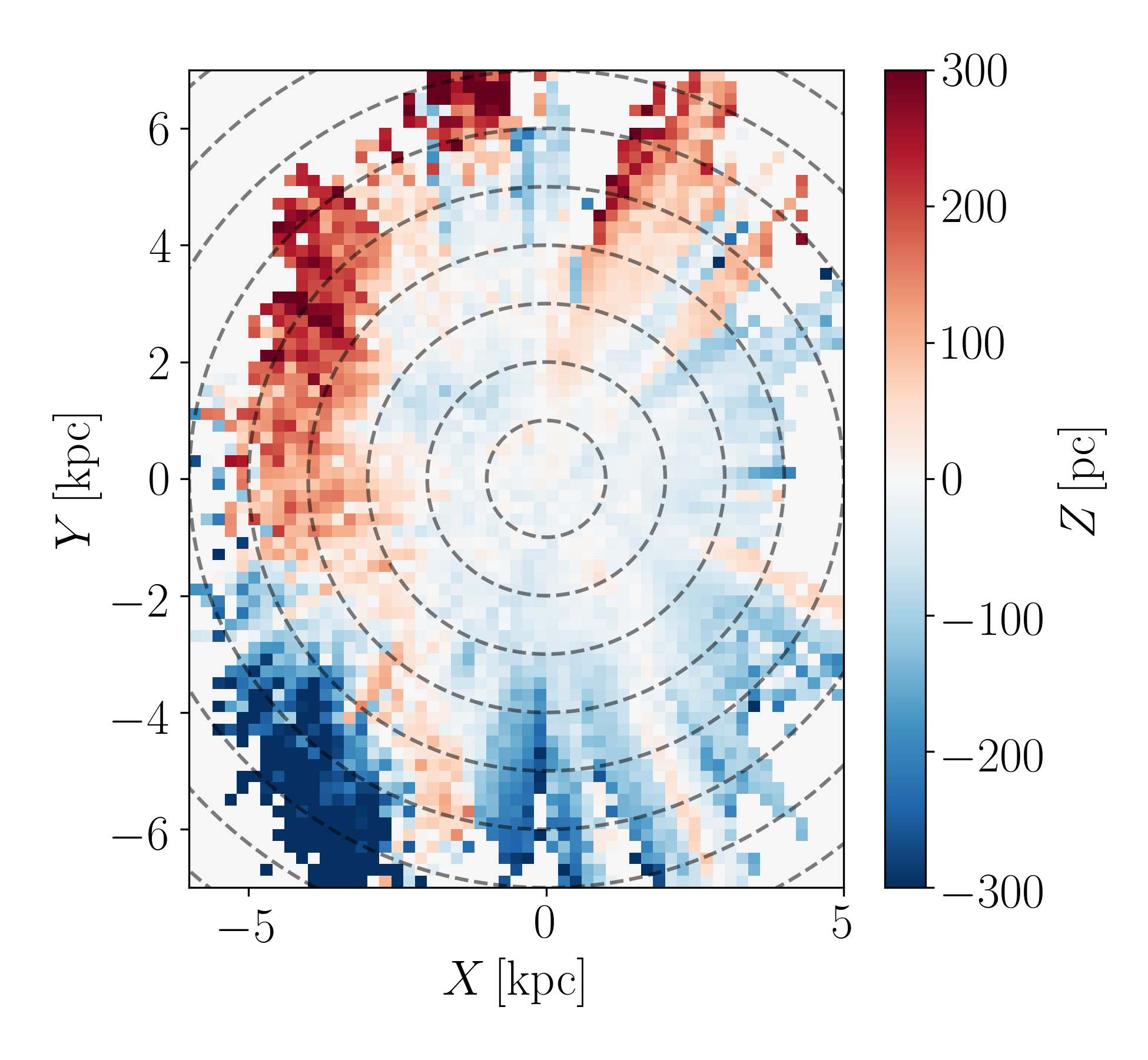}
    \caption{Median height for the filtered sample projected on the Galactic plane. We divided the XY plane into bins of 200 pc width,  and we only show the ones containing more than 10 stars. The dashed circles have radii from 1 to 9 kpc, in steps of 1 kpc. The radial features are likely due to uneven sampling of sources in the Galactic plane due to foreground extinction.}
    \label{fig:median_z}
\end{figure}

Fig. \ref{fig:median_z} shows a map of the median height $z$ of our sample on a spatially uniform grid, with bins of 100 pc width. We consider only bins that have at least 10 stars.  
The radial features in Fig. \ref{fig:median_z} are likely an artefact due to uneven sampling above or below the Galactic plane due to foreground extinction. Such features make the interpretation of Fig. \ref{fig:median_z} uncertain and might prevent from seeing the warp shape, especially for $X > -2.5$ kpc.
For $X < -2.5$ kpc the impact of extinction is less dramatic (see also Fig. \ref{fig:dust_and_stars}, thus the displacement visible above and below the plane might indeed trace the warp \citep[see also Fig. 5 in][]{Romero2019}.

\begin{figure}
    \centering
    \includegraphics[width = \hsize]{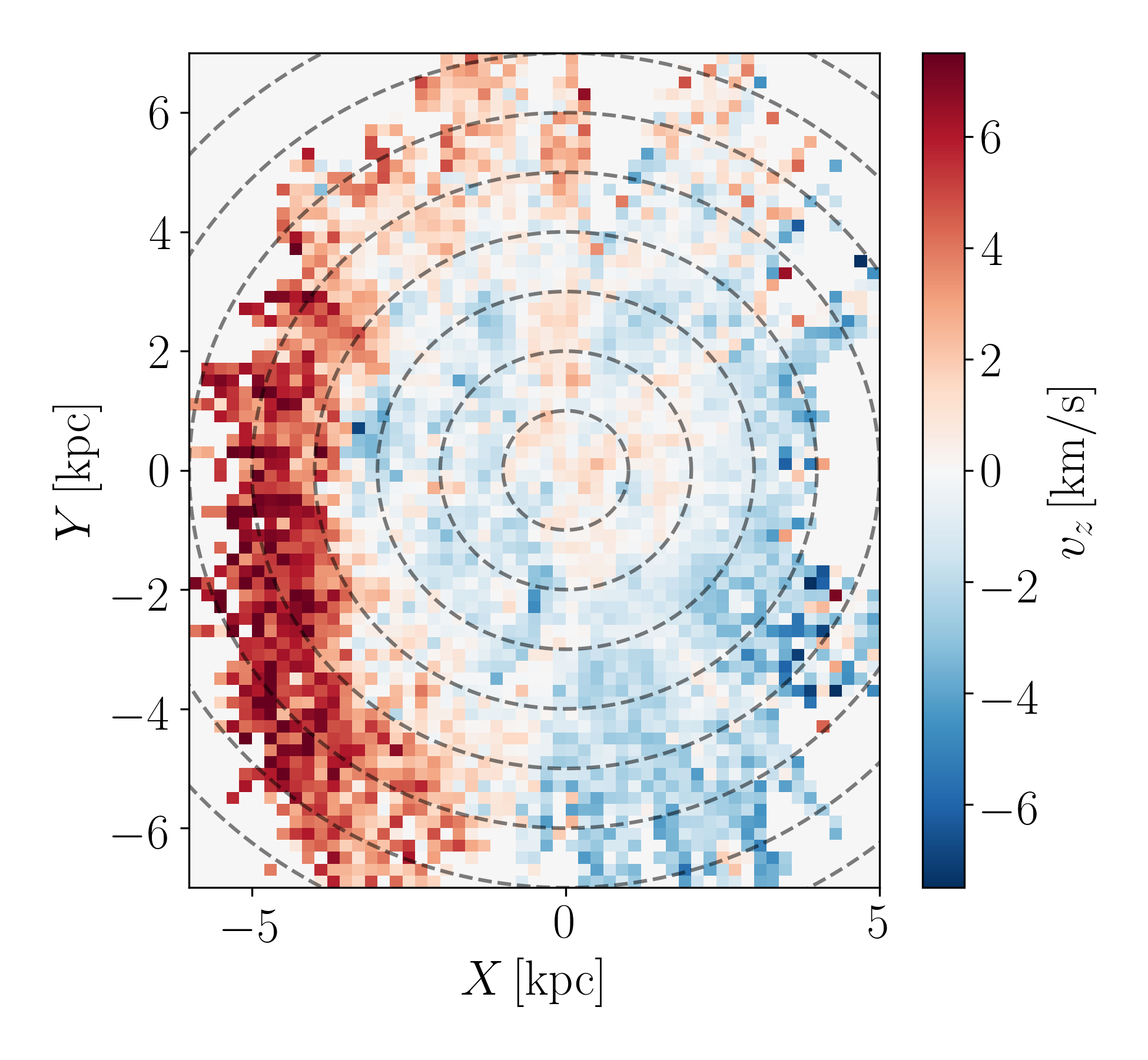}
     \caption{Median vertical velocity $v_z$ distribution on the Galactic plane. We divided the XY plane into bins of 200 pc width,  and we only show the ones containing more than 10 stars. The dashed circles have radii from 1 to 9 kpc, in steps of 1 kpc. The Galactic warp is visible for $R > 4$ kpc towards the Galactic anti-centre.}
    \label{fig:median_vz}
\end{figure}

The kinematic signature of the warp is evident in Fig. \ref{fig:median_vz} \citep[see e.g.][]{Poggio2018, Romero2019}, which shows the median vertical velocity ($v_z$) distribution of our filtered sample on the Galactic plane (see Appendix \ref{sec:vertical velocity} for more details on how the vertical velocity was computed). For distances larger than 4 kpc from the Sun, the vertical velocities have positive values that seem to peak at $Y \sim 0$ kpc and decrease towards both sides. At smaller distances from the Sun, some features also present positive vertical velocity. While some correspond to density enhancements in the 3D spatial distribution of our sources, this is not the case for all of them. Such variation in $v_z$ are likely related to the perturbations that the disc is experiencing (e.g. interaction with satellite galaxies).

\subsection{Comparison with other O and B star maps}\label{sec:ob}
By combining \textit{Gaia} EDR3 and VPHAS data with literature catalogues, \cite{Chen2019} obtained a sample of  14 880 OB stars, earlier than B3V, which they used to describe the morphology of the spiral structure of the Milky Way. Chen et al. method to select O and B stars substantially differs from ours, however  a qualitative comparison between their Fig. 4 and our maps shows substantially the same gross structures. This increases our confidence that both maps are revealing the same features in the Galactic O and B star distribution. The larger number of sources in our catalogue enables us to reveal more details in the source distribution. Interestingly, neither maps show a clear spiral structure. This will be further discussed in the Sec. \ref{sec:discussion}.

Fig. 3 in \cite{Romero2019} shows the distribution on the Galactic plane of their upper main sequence sources. Fig. \ref{fig:map0} shows essentially the same over-densities, although to a higher level of detail. This is mainly due to two facts: a) to create their maps, \cite{Romero2019} used a larger bin size than in our present work, and b) their sample likely includes more later type stars than ours due to their selection criteria: this causes the distribution of sources to be smoother. 

Fig. 3 (panel A) in \cite{Poggio2018} also shows the distribution of upper main sequence sources on the Galactic plane. Similarly to \cite{Romero2019}, the pixel size of \cite{Poggio2018} is higher than ours. As mentioned above, this creates a smoother map, where it is not possible to identify small-scale over-densities. The comparison between large scale structures shows however many similarities.

\subsection{Comparison with dust distribution} \label{sec:dust_maps}
\begin{figure*}
    \centering
    \includegraphics[width = 1.1\hsize]{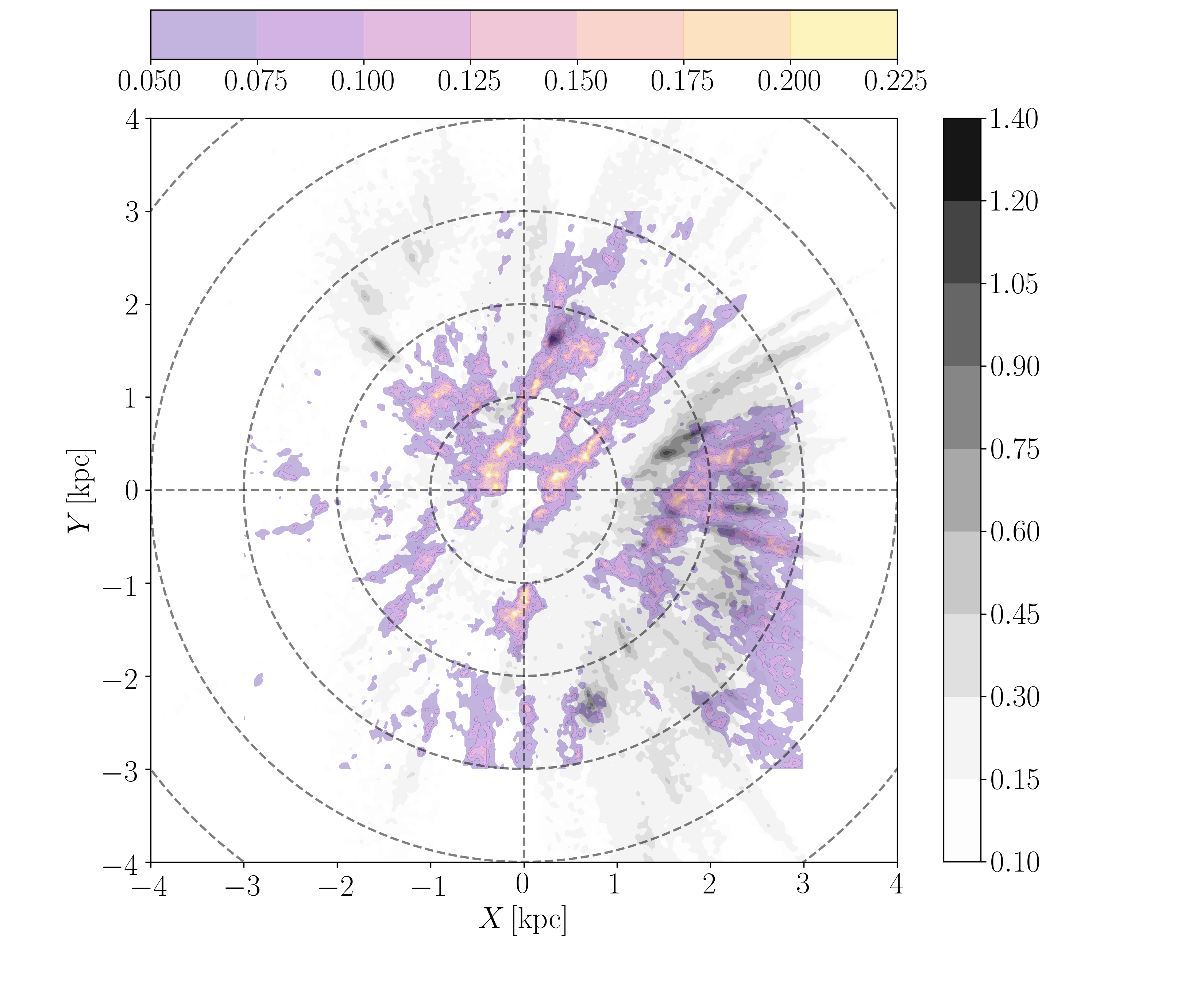}
    \caption{Same as Fig. \ref{fig:map0}, with dust density contours from \cite{Lallement2019} over-plotted. Both densities are displayed in arbitrary units. The dashed circles have radii from 1 to 5 kpc, in steps of 1 kpc.}
    \label{fig:dust_and_stars}
\end{figure*}

Dust is also a  tracer of Galactic structure. 
Fig. \ref{fig:dust_and_stars} shows the projection of the 3D dust distribution from \cite{Lallement2019} on the Galactic plane, over-plotted on top of our density map (see Fig. \ref{fig:map0}).  The units  of the density distribution are arbitrary. The dust density distribution shows discrepancies with respect to the star distribution. For example, the two elongated structures at the centre of the map are not prominent in the OB stars distribution, and the dust features in the first and fourth Galactic quadrants are off-set with respect to the star density distribution, and seem to have a different separation and relative inclination.  Such offsets are expected, as newly formed stars will drift from their birthplaces as they follow galactic rotation while the spiral arms move with a given pattern speed. 

\subsection{Comparison with Cepheid distribution}\label{sec:cepheids}
\begin{figure*}
    \centering
    \includegraphics[width = \hsize]{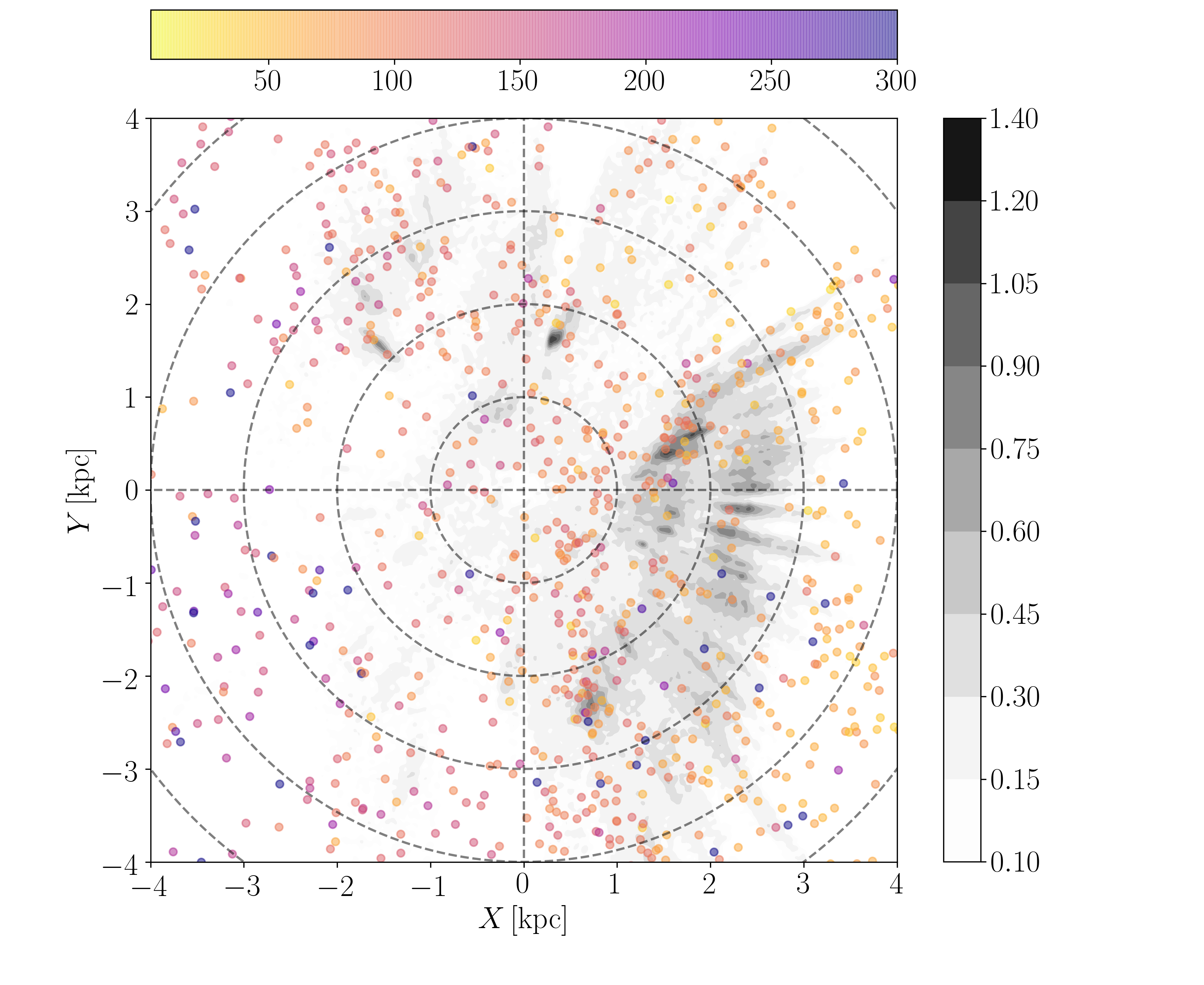}
    \caption{Cepheids from \cite{Skowron2019}, colour-coded by age (Myr). The density map is the same as Fig. \ref{fig:map0}. The dashed circles have radii from 1 to 6 kpc, in steps of 1 kpc.}
    \label{fig:cepheids}
\end{figure*}

Classical Cepheids are young (< 400 Myr) variale stars, whose distances can be estimated thanks to their period-luminosity relation. (Fig. \ref{fig:cepheids} show the Cepheids identified by \cite{Skowron2019}, over-plotted on the density distribution of O and B stars (same as in Figs. \ref{fig:map0} and \ref{fig:dust_and_stars}). The colour-bar represents the ages determined by \cite{Skowron2019} (in million years). The Cepheid distribution traces reasonably well the density enhancements corresponding to the Sagittarius-Carina arm (see Sec. \ref{sec:reid}), while the correspondence with the other density enhancements is not as tight. Although the age distribution of the Cepheids and our filtered sample is similar, the selection criteria are different: this makes a more direct comparison difficult.

\subsection{Comparison with maser distribution}\label{sec:reid}
\begin{figure*}
    \centering
    \includegraphics[width = \hsize]{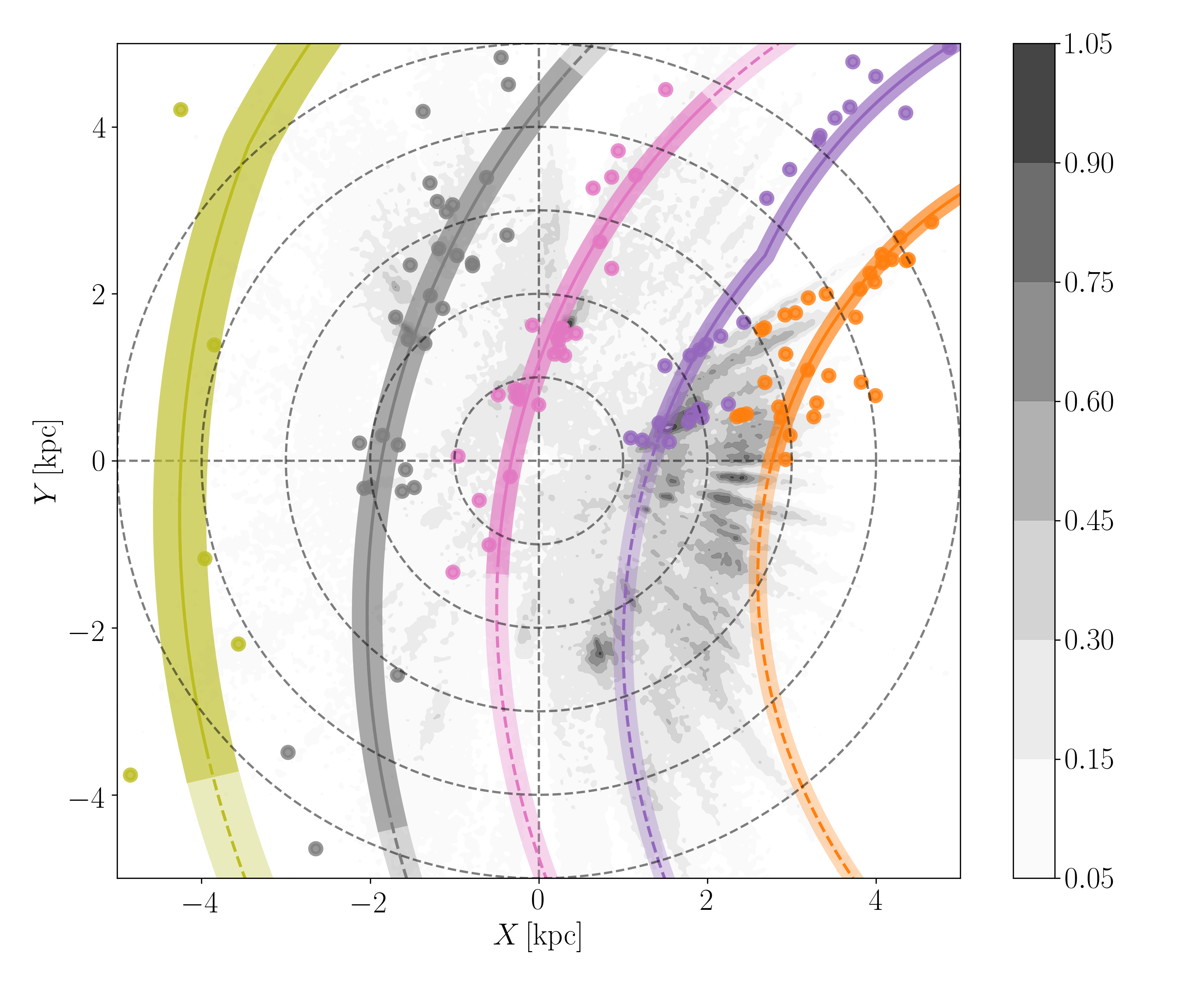}
    \caption{The masers (dots) and the fit to the spiral arms (solid lines) from  \cite{Reid2019} are over-plotted on top of our density map (same as Fig. \ref{fig:map0}). The maser distances were computed by naively inverting the parallax. The dashed circles have radii from 1 to 6 kpc, in steps of 1 kpc.
    The shaded regions correspond to the arm width. Different colours correspond to different spiral arms: Outer, \textit{green}; Perseus, \textit{grey}; Local, \textit{pink}; Sagittarius-Carina, \textit{purple}; Scutum-Centaurus, \textit{orange}. }
    \label{fig:spiral}
\end{figure*}

The spiral structure of the Milky Way can be traced also by water and methanol masers associated with high-mass star forming regions (HMSFRs). By measuring parallaxes and proper motions of such HMSFRs, \cite{Reid2019} provided estimates of the fundamental Galactic parameters, among which the pitch angle and the arm width. Fig. \ref{fig:spiral} shows the masers and the location of the arms from the fit from \cite{Reid2019}, on top of the OB stars density map shown in Fig. \ref{fig:map0}.
Similarly to the Cepheids distribution, there is a good agreement between the position of the masers and the location of the most prominent over-densities, which makes us confident of our distance estimates. Unfortunately there are no available maser data tracing the spiral structure in the 4th Galactic quadrant ($X > 0$, $Y < 0$), thus making a more detailed comparison in that region non-trivial. However, by tracing the spiral arms beyond the data while keeping the parameters fixed (dashed lines, and lighter contours), more discrepancies become visible, especially towards the inner Galaxy, towards the Sagittarius-Carina (purple) and  Scutum-Centaurus arm (orange).

\section{Discussion}\label{sec:discussion}
We have devised a large and systematically selected sample of massive, young stars and we studied its properties focusing on the spatial distribution of our sources on the Galactic plane. In this Section we discuss our findings in the context of the spiral arm structure of the Milky Way.

Traditionally, the four main spiral arms of the Milky Way are considered to be Perseus, Sagittarius, Scutum and Norma, with an additional Outer arm, which might be the outer part of one of the inner arms, and the Local or Orion arm, which is shorter and may be a spur or a bridge between two arms \citep[see for instance][]{Churchwell2009}. In the Milky Way,  the terminology "spiral arm" has been used very broadly, for HI, molecular gas (and masers), dust and young stars, while it has been shown from external galaxies that these tracers have quite distinct morphology. This has led to some inconsistencies in defining the spiral arms in our Galaxy.
For example, the structure parameters derived by \cite{Reid2019} (see Fig. \ref{fig:spiral} and Sec. \ref{sec:reid}) do not agree with those presented in \cite{Chen2019} (see Sec. \ref{sec:ob}).
Further, certain features might be interpreted as spiral arms or spurs connecting different arms: this is for example the case for the dust complex labelled as "Lower Sagittarius-Carina" by \cite{Lallement2019} (see Sec \ref{sec:dust_maps} and Fig. \ref{fig:dust_and_stars}) and "Lower Sagittarius-Carina Spur" by \cite{Chen2020}. 
Further, by studying the correlation between the location of young clusters and molecular clouds in NGC 7793 and M51 (respectively a flocculent and a grand-desing spiral galaxy), \cite{Grasha2018} and \cite{Grasha2019} found that the star clusters that are associated (i.e., located within the footprint of a giant molecular cloud) are young, with a median age of 2 Myr in NGC 7793 and of 4 Myr in M51. Older clusters are mostly un-associated with any molecular cloud. Thus, equating the same spiral arm morphology with different tracers (such as early type stars and dust) might be inappropriate. 

\begin{figure*}    
    \centering
    \includegraphics[width = 0.33\hsize]{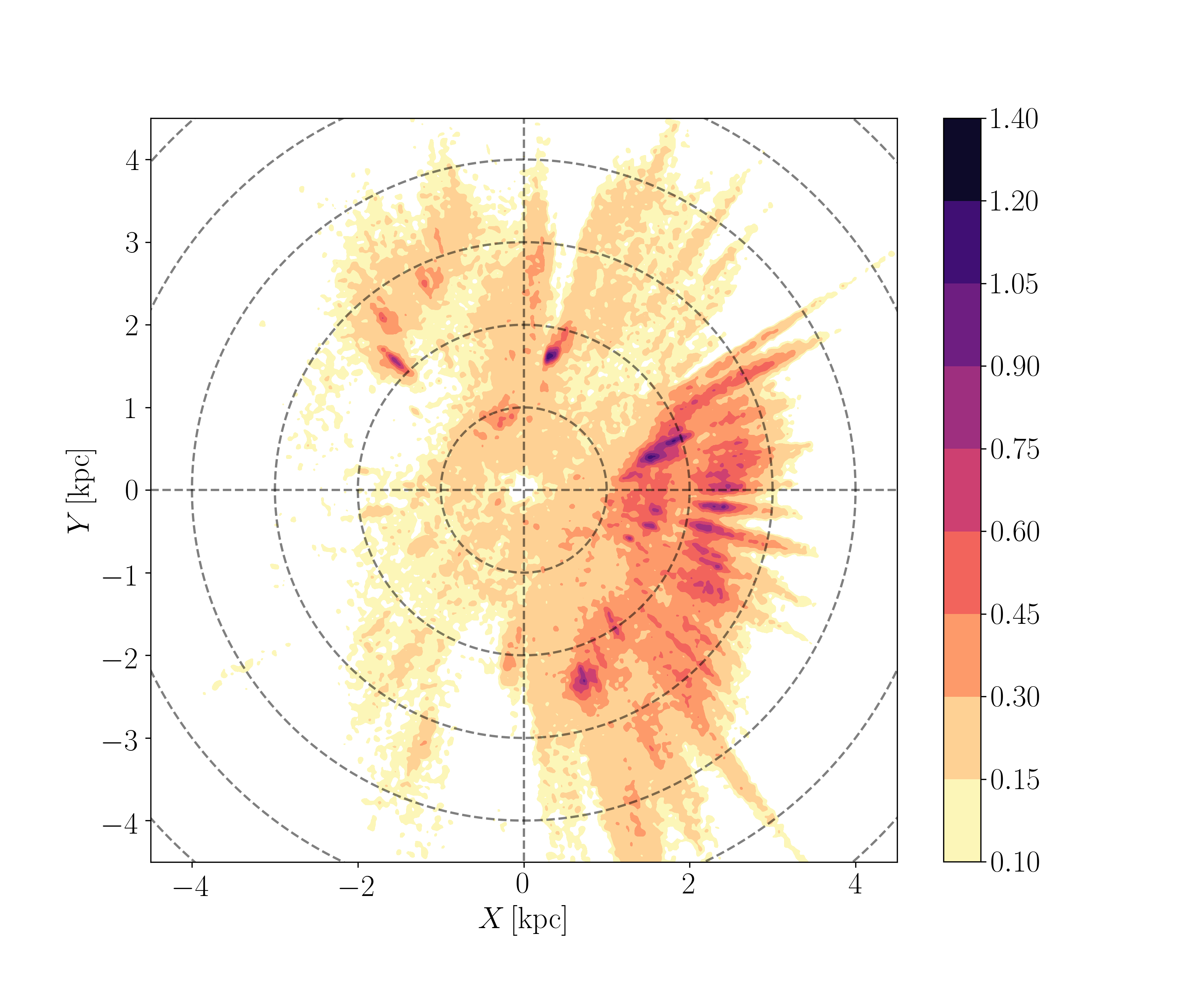}
    \includegraphics[width = 0.33\hsize]{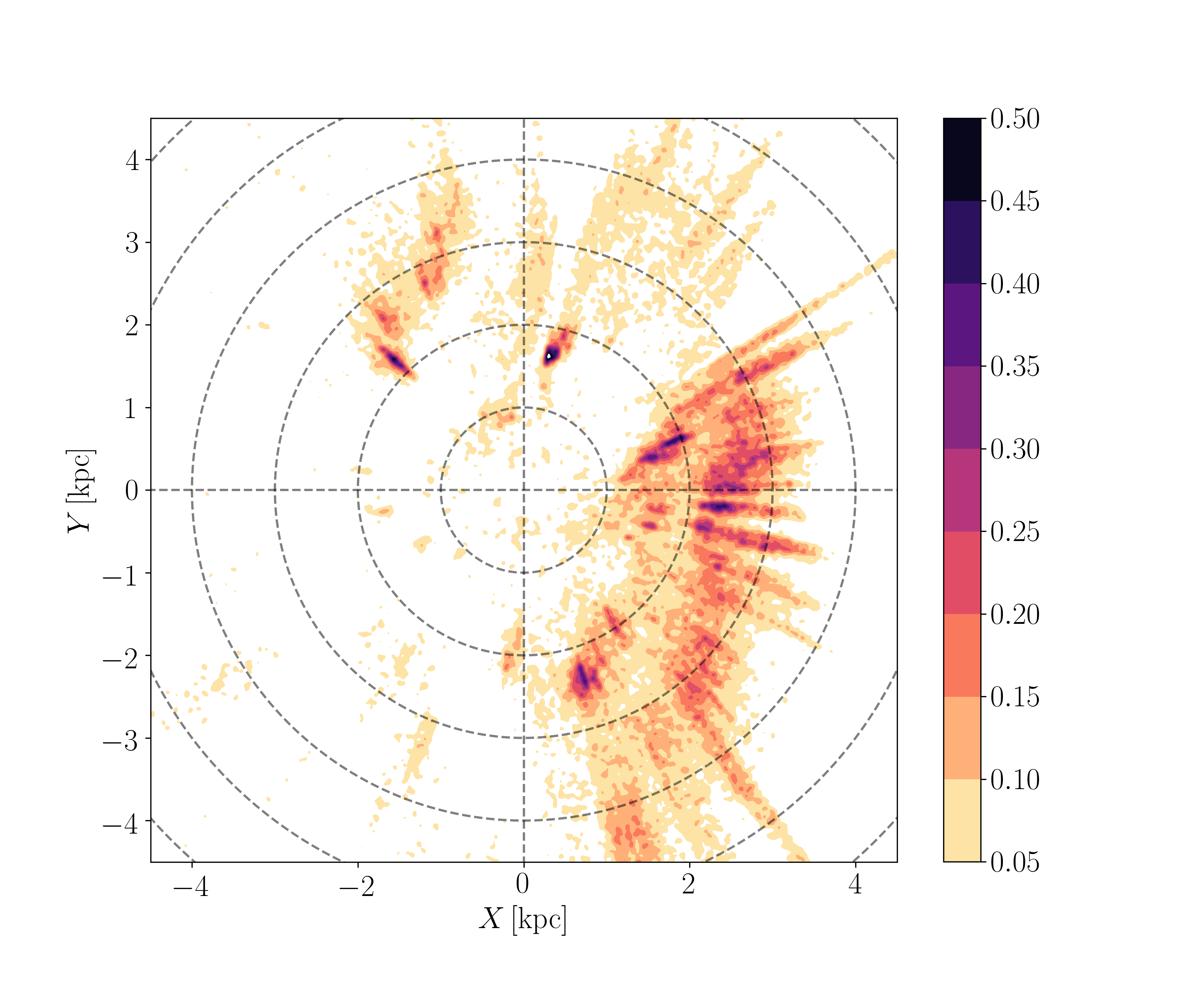}
    \includegraphics[width = 0.33\hsize]{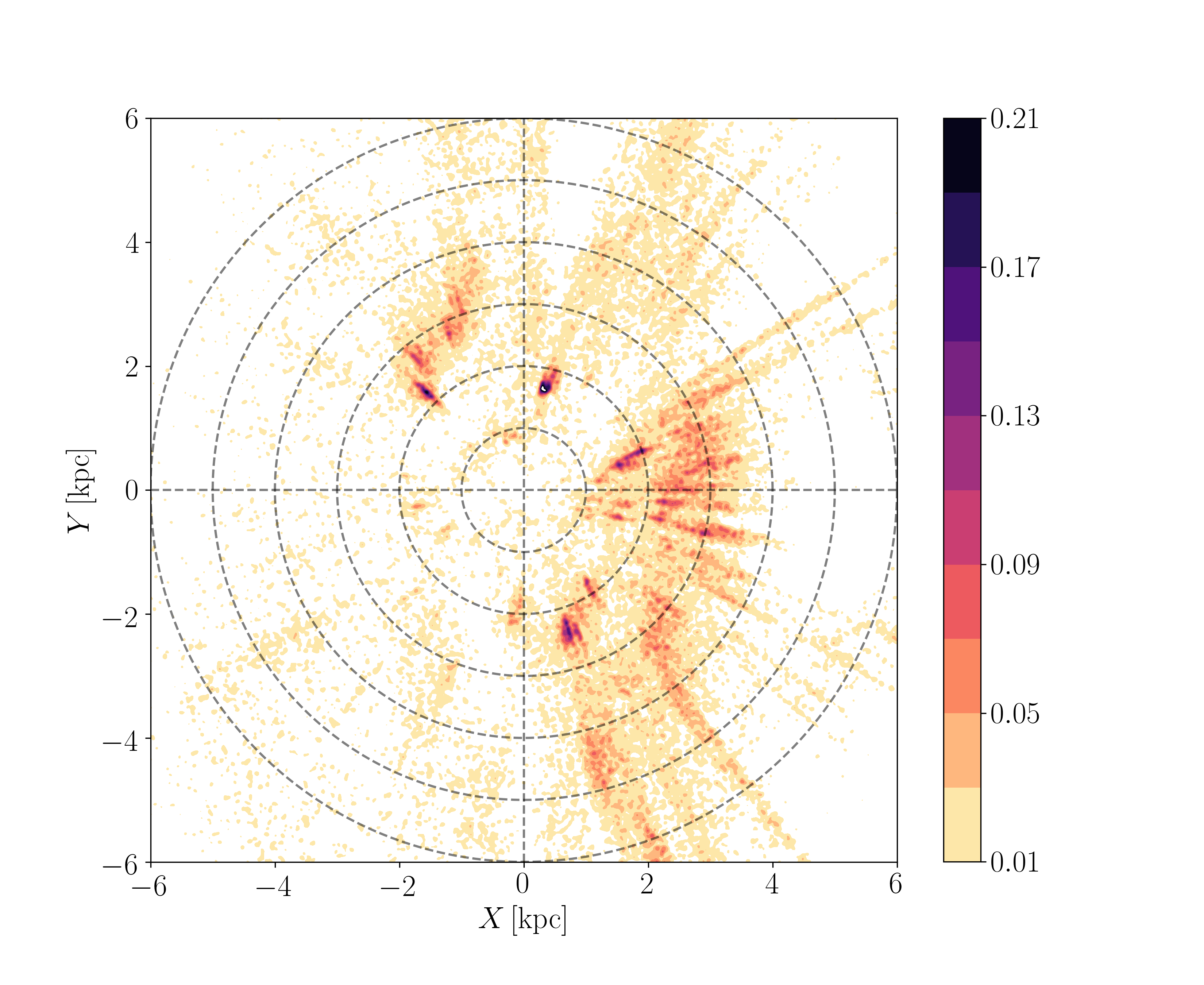}
    \caption{Left: same as Fig. \ref{fig:map0}. Centre: density of stars with $M_{K_s} < -1$ mag projected on the Galactic plane. Right: density of stars with $M_{K_s} < -2$ mag projected on the Galactic plane. Note that the density scales are different.}

    \label{fig:map1}
\end{figure*}

The over-density towards the inner Galaxy (visible for positive $x$ values, i.e. in the $1st$ and $4th$ Galactic quadrant) in Fig. \ref{fig:map0} is associated with the Sagittarius-Carina and Scutum-Centaurus arm, and is much more prominent than the others, containing numerous high-mass star forming regions.
On the contrary, the distribution of massive stars associated with the Perseus arm peaks mainly in the $2nd$ Galactic quadrant (towards Cassiopeia). This is consistent with \cite{Reid2019} findings, and would point to the conclusion that the Perseus arm as traced by O and B-type stars is not a dominant arm, and might be dispersing in the field.
This suggests that a recent, large-scale episode of (massive) star formation occurred in Sagittarius-Carina and Scutum-Centaurus, while in the other arms (Perseus and Local) stars formed earlier, except in a few isolated massive associations, such as Cygnus and Cassiopeia.  
To confirm this,  we selected stars brighter than $M_{K_s} = -1 \, \mathrm{mag}$ and $M_{K_s} = -2 \, \mathrm{mag}$
(corresponding respectively to the $K_s$ absolute magnitude of  B2V-type stars and B1V-type star), and we evaluated their density on the plane following the procedure described in Sec. \ref{sec:results}. The density maps obtained with these samples are shown in Fig. \ref{fig:map1}, where also the same map of Fig. \ref{fig:map0} is shown for clarity. The density contours in  Fig. \ref{fig:map1} (left) trace the same dense structures as in Fig. \ref{fig:map0} (note that the contour levels are different), while the low-density contours substantially disappear. This is even more evident in Fig. \ref{fig:map1} (right). This is expected, as intrinsically brighter stars are on average younger and thus have had less time to disperse in the field.

\cite{Drimmel2000b} and \cite{Drimmel2001} studied the spiral arm features of the Milky Way in the far-infrared and near-infrared and concluded that the near-infrared features were consistent with a two-armed spiral model, while the other two arms were traced only by the far-infrared emission and thus could perhaps be present only in gas and young stars. 
More recently, \cite{Xu2018} and \cite{Chen2019} noted that the spiral arm structure traced by  O and early B-type stars may have many sub-structures in addition to the four major arms. Thus, they concluded that the Milky Way might not be pure grand design spiral galaxy with well-defined, two or four dominant arms. This might suggest that the Milky Way could exhibit characteristics of both flocculent and grand design arms, with grand design seen in the infrared (old stellar populations) and more flocculent (or multi- arm) features seen in the optical (young stars). 
As mentioned above, these characteristics have been observed in spiral galaxies for instance by \citet[and references therein]{Kendall2011, Kendall2015}, who concluded that galaxies that exhibit a grand design structure in the optical also exhibit such structure in the NIR, while there are optically flocculent galaxies which do exhibit NIR grand design structure. As pointed out by \cite{Dobbs2014},
\cite{Elmegreen1999} and \cite{Elmegreen2011} noted however that most flocculent galaxies do not exhibit grand-design structure and those that do have very weak spiral arms. 
\cite{Elmegreen2003} further suggested that both grand-design and flocculent spirals (as seen in the old stars) exhibit a similar structure in the gas and young stars (independent of the underlying old stellar population) which is driven by turbulence in the disc.
The co-existence of different spiral arm patterns in the Milky Way disc might have different explanations. For example,  
\cite{Martos2004} concluded that secondary arms could be resonance-related features in the quasi-stationary density wave picture, while \cite{Steiman-Cameron2010} and \cite{Drimmel2000b} proposed that the  traditional four spiral arms may represent the dynamical response of a gas disc to a two-arm spiral perturbation in the mass distribution. 

As mentioned before, the density distribution presented in Fig. \ref{fig:map0} exhibits a prominent density enhancement roughly corresponding to the Sagittarius-Carina and Scutum-Centaurus arms. This alone does not allow to draw firm conclusions on the nature (nor on the number) of the spiral arms of the Milky Way. On the one hand, star formation occurs in a clumpy and patchy fashion: this might explain the observed density distribution without necessarily assuming a flocculent structure.
On the other hand, the map shown in Fig. \ref{fig:cepheids} might suggest a different picture, in which the Milky Way, as traced by young stars ($\lesssim 300$~Myr) does not show a set of discrete and distinct set of spiral arms.

The spiral arm structure of the Milky Way can also be investigated by using stellar kinematics \citep[see e.g.][]{Antoja2016}. Of course, a spiral mass density perturbation can have a very different morphology from spiral-like density enhancements in molecular gas or young stars.
\cite{Eilers2020} studied the kinematics of red giant stars and showed a spiral feature in their radial velocities that they interpret as rising from a perturbation to the potential of the Milky Way caused by a two-armed logarithmic spiral. Their model predicts the locations of the spiral perturbation in the Milky Way, one roughly co-spatial to the Orion (or Local) arm, and the other roughly corresponding to the Outer arm. The location of the perturbation predicted by \cite{Eilers2020} does not closely follow the over-densities of the map in Fig. \ref{fig:map0}. The comparison between the radial velocities of the O and B-type star sample and the giant sample will be crucial to make progress in our understanding of the mechanism giving rise to the spiral signature presented by \cite{Eilers2020}. The data of the SDSS-V survey will make such a comparison possible, and will also likely allow to construct a surface density map of the giant sample which could be directly compared with the map presented in this work.
The complete kinematic information that we will obtain by combining \textit{Gaia} proper motions with SDSS-V radial velocities will also allow to better separate stars belonging to different spiral arms, and to study their internal motions. Indeed, as already mentioned in Section \ref{sec:results}, vertical velocities are not related to the perturbation in the disc kinematics induced by the spiral arms.

\section{Conclusions}\label{sec:conclusion}
In this study we have analysed the 3D space distribution of a sample of hot and luminous OBA stars, focusing in particular on their configuration in the Galactic disc.

Our target selection is based on the combination of \textit{Gaia} EDR3 astrometry and photometry and 2MASS photometry, and is aimed at providing a well-defined selection function. 
We describe the properties of the target sample in terms of distribution in the sky, brightness, and variability. We estimate the purity and contamination of the catalogue by comparing with existing O and B-type star catalogues, such as those by \cite{Liu2019}, and  \cite{Sota2014} and \cite{Maiz2016}, and by deriving stellar parameters ($T_{\mathrm{eff}}$ and $\log g$) for stars in common with LAMOST DR6.

By assuming that young massive stars are on near-circular orbits with a small velocity dispersion, we compute astro-kinematic distances, and compare them to those derived by \cite{Bailer-Jones2020}. We use these distances to study the distribution of sources in the Galactic disc of a sub-set of the target sample, which we obtain by filtering out sources with spurious astrometric solutions and kinematic properties inconsistent with our model.

We find that the distribution of sources in the plane of the Milky Way is highly structured, characterised by over- and under-densities. 
Some of the density enhancements correspond to massive star forming regions, such as Carina, Cygnus, and Cassiopeia. With these associations, the inner arms (Sagittarius-Carina and Scutum-Centaurus) are strikingly more prominent in OBA stars than any outer arms, for which we find little evidence.

The distribution of O and B-type stars from previous catalogues, classical Cepheids, dust,  and high-mass star forming regions shows similarities with our density distribution. However, the picture of the spiral arm structure of the Galaxy that we obtain in this study is complex and it may suggest that young stars show little tendency to be neatly organised in distinct spiral features. 

To assess different spiral arm models, it would be necessary to extend our map beyond the volume currently studied and, perhaps more importantly, complement with better \textit{Gaia} data and spectroscopic information.

The target sample that we have devised is indeed optimised for spectroscopic follow-up with the SDSS-V survey.
The information that we will be able to obtain with SDSS-V will be  crucial for studying the kinematics and dynamics of the spiral arms, and thus the nature of the spiral arms themselves. Finally, the sample that we have devised will further allow to study the properties of massive stars, for example in terms of multiplicity, internal structure, and (binary) evolution, in a statistic fashion. 

\begin{acknowledgements}
We thank the referee for their comments, which improved the quality of this manuscript.
We would also like to thank: J. Rybizki and G. Green for making the catalogue of spurious sources available in advance of publication; A. Gould for discussions that prompted us to use kinematics for distance estimations; M. Sormani for discussions on the kinematics of the MW bar and bulge.\\
This work has made use of data from the European Space Agency (ESA) mission
{\it Gaia} (\url{https://www.cosmos.esa.int/gaia}), processed by the {\it Gaia}
Data Processing and Analysis Consortium (DPAC,
\url{https://www.cosmos.esa.int/web/gaia/dpac/consortium}). Funding for the DPAC
has been provided by national institutions, in particular the institutions
participating in the {\it Gaia} Multilateral Agreement.\\
This publication makes use of data products from the Two Micron All Sky Survey, which is a joint project of the University of Massachusetts and the Infrared Processing and Analysis Center/California Institute of Technology, funded by the National Aeronautics and Space Administration and the National Science Foundation.\\
Funding for the Sloan Digital Sky Survey V has been provided by the Alfred P. Sloan Foundation, the Heising-Simons Foundation, and the Participating Institutions. SDSS acknowledges support and resources from the Center for High-Performance Computing at the University of Utah. The SDSS web site is www.sdss.org.
SDSS is managed by the Astrophysical Research Consortium for the Participating Institutions of the SDSS Collaboration, including the Carnegie Institution for Science, the Chilean Participation Group, the Gotham Participation Group, Harvard University, The Johns Hopkins University, L'Ecole polytechnique fédérale de Lausanne (EPFL), Leibniz-Institut für Astrophysik Potsdam (AIP), Max-Planck-Institut für Astronomie (MPIA Heidelberg), Max-Planck-Institut für Extraterrestrische Physik (MPE), Nanjing University, National Astronomical Observatories of China (NAOC), New Mexico State University, The Ohio State University, Pennsylvania State University, Space Telescope Science Institute (STScI), the Stellar Astrophysics Participation Group, Universidad Nacional Autónoma de México, University of Arizona, University of Colorado Boulder, University of Illinois at Urbana-Champaign, University of Toronto, University of Utah, University of Virginia, and Yale University.\\
The research leading to these results has (partially) received funding from the KU~Leuven Research Council (grant C16/18/005: PARADISE), from the Research Foundation Flanders (FWO) under grant agreement G0H5416N (ERC Runner Up Project), as well as from the BELgian federal Science Policy Office (BELSPO) through PRODEX grant PLATO.\\
This research made use of TOPCAT \citep{Topcat}, Astropy, \citep{Astropy2013, Astropy2018}, matplotlib \citep{matplotlib}, numpy \citep{harris2020array}, scipy \citep{2020SciPy-NMeth}, and scikit-learn \citep{scikit}. This work would have not been possible without the countless hours put in by members of the open-source community all around the world.

\end{acknowledgements}

\bibliographystyle{aa} 
\bibliography{bibliography}

\appendix
\section{Query of the \textit{Gaia} archive}\label{sec:query}
Here we provide an example query for cross-matching 2MASS and \textit{Gaia} EDR3 in the \textit{Gaia} archive by using the cross-match with \textit{Gaia} DR2. 

\texttt{select edr3.*
xdr2.*, tm.* \\
from gaiaedr3.gaia\_source as edr3 \\
    inner join gaiaedr3.dr2\_neighbourhood as xdr2  \\
        on edr3.source\_id = xdr2.dr3\_source\_id \\
    inner join gaiadr2.tmass\_best\_neighbour as xtm \\ 
        on xdr2.dr2\_source\_id = xtm.source\_id \\
	inner join gaiadr1.tmass\_original\_valid AS tm \\
		on tm.tmass\_oid = xtm.tmass\_oid \\
WHERE  xtm.angular\_distance < 1.  \\
AND xdr2.angular\_distance < 100. \\
AND edr3.phot\_g\_mean\_mag < 16. \\
AND edr3.parallax $<$ power(10,(10-tm.ks\_m)/5)}

\section{Astro-kinematic distances}\label{sec:kin_dist}
In this Section we describe all the terms of Eq. \ref{eq:pdf_text}, which we report here for convenience:
\begin{align}
    &p(d~|~ \Vec{o}, m_{K_s} \Theta_{\mathrm{KM}}, \Theta_{\mathrm{SM}}, \Theta_{\mathrm{CMD}})  \propto \nonumber \\ 
    &p(\Vec{o} ~|~ \Theta_{KM}) p(d, m_{K_s} ~|~ l, b, \Theta_{\mathrm{SM}}, \Theta_{\mathrm{CMD}}).
\label{eq:pdf} 
\end{align}
The prior $p(d, m_{K_s} ~|~ l, b, \Theta_{\mathrm{SM}}, \Theta_{\mathrm{CMD}})$ can be written as:
\begin{equation}
    p(d, m_{K_s} ~|~ l, b, \Theta_{\mathrm{SM}}, \Theta_{\mathrm{CMD}}) \propto p(d ~|~ l, b, \Theta_{\mathrm{SM}}) ~ f (d, m_{K_s} ~|~ \Theta_{\mathrm{CMD}}).
\end{equation}
The term $p(d | l, b, \Theta_{\mathrm{SM}})$ represents the probability density function of observing a star in the direction $(l,b)$ at distance $d$ from the Sun according to our assumed model for the spatial distribution of stars in the Galaxy. The term  $f (d, m_{K_s} | \Theta_{\mathrm{CMD}})$ specifies the fraction of stars that can be observed at a distance $d$ and magnitude $m_{K_s}$ given the distribution of stars in colour-magnitude space and our selection criteria.
In the following Sections, we describe the different components of Eq. B.1.

\begin{figure*}
    \centering
    \includegraphics[width = 0.45\hsize]{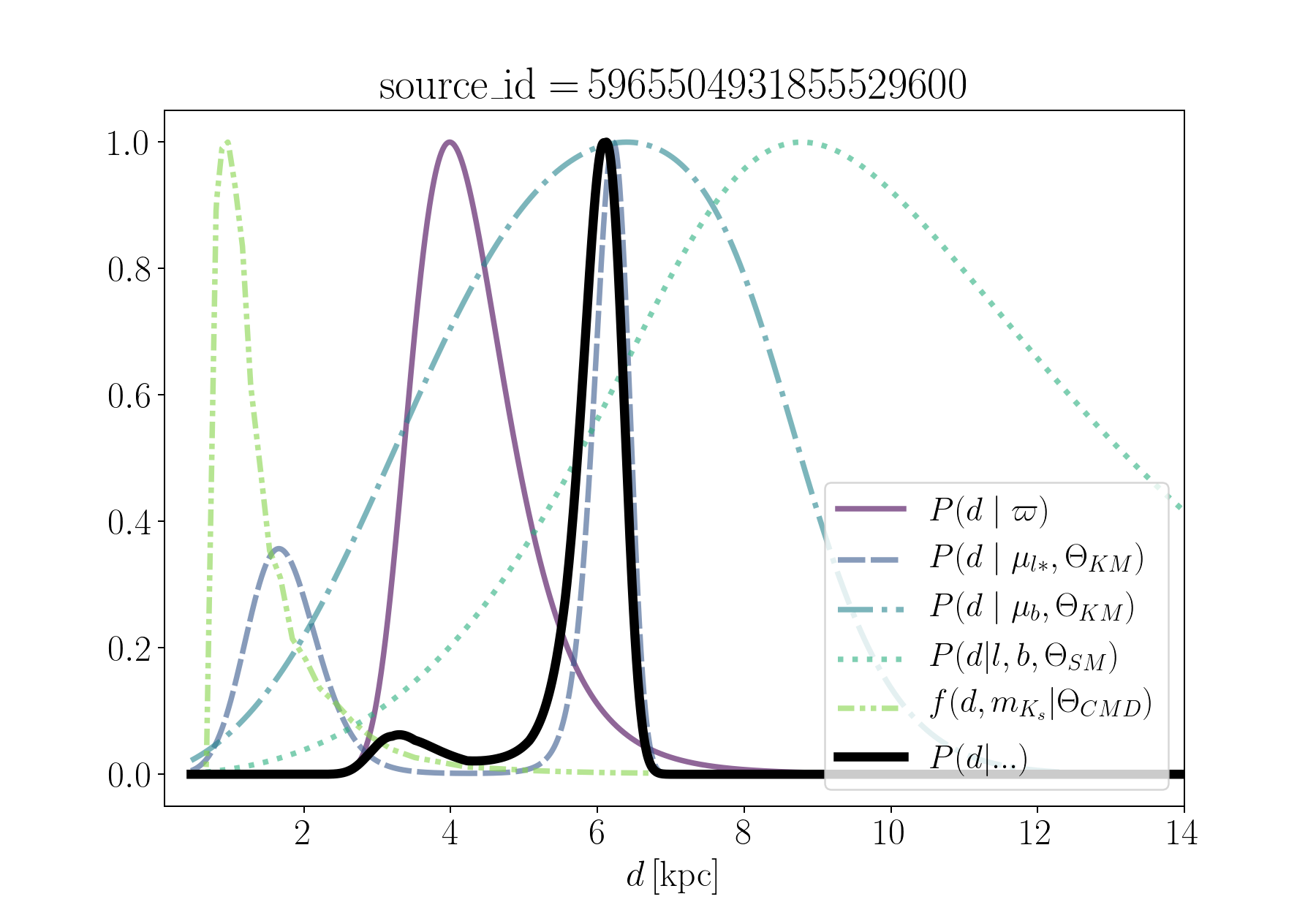}
    \includegraphics[width = 0.45\hsize]{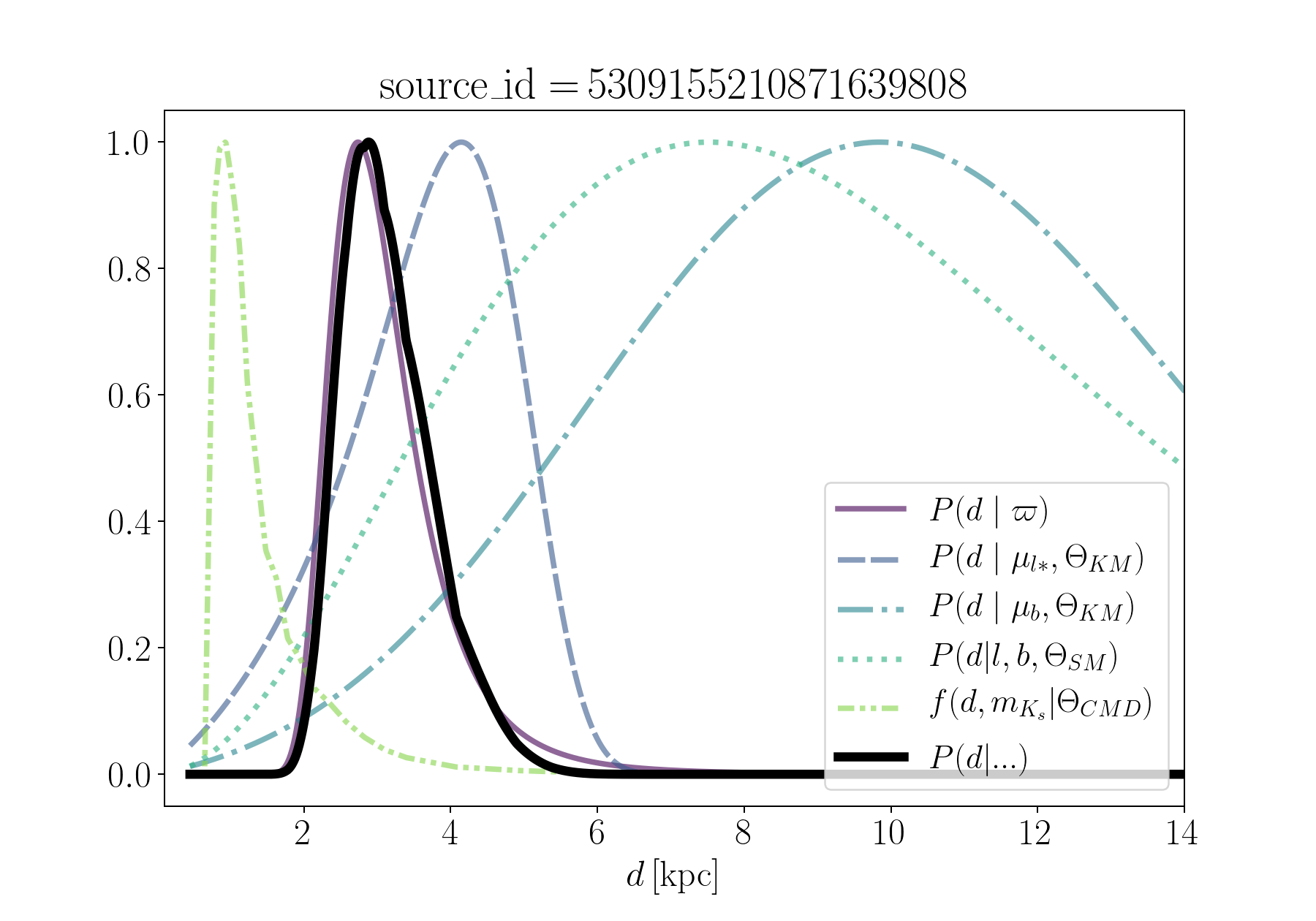}
    \includegraphics[width = 0.45\hsize]{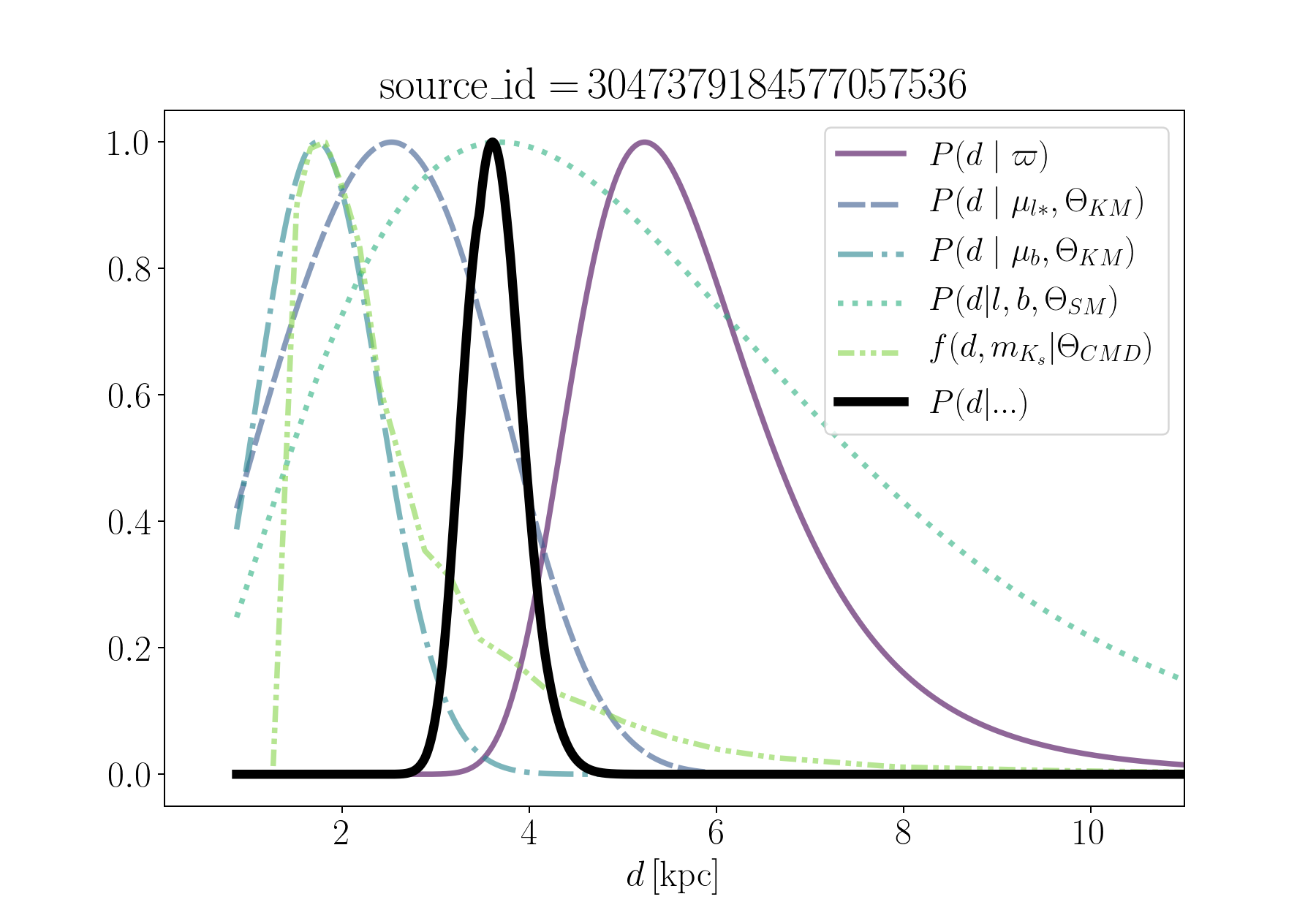}
    \includegraphics[width = 0.45\hsize]{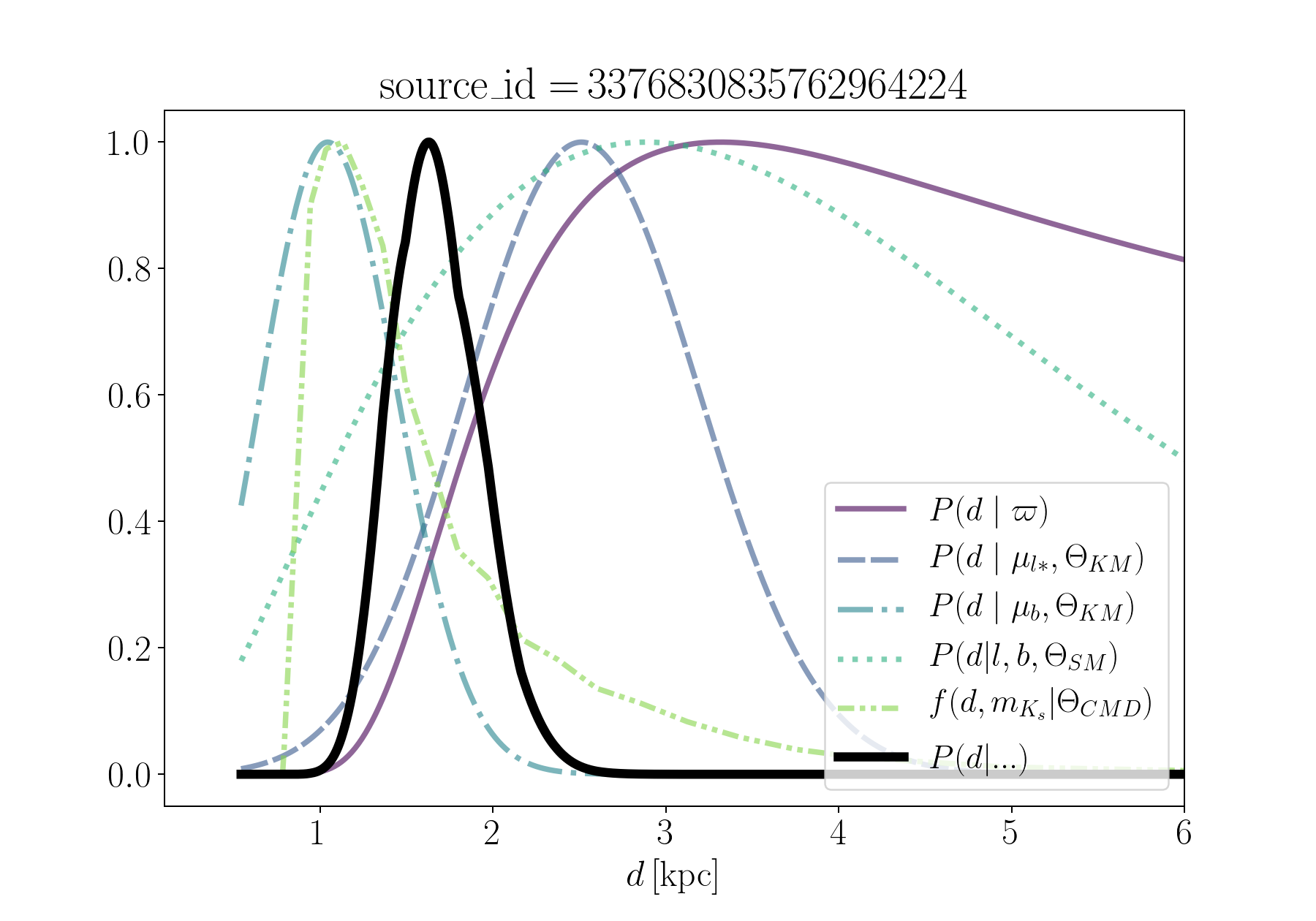}
    \caption{Posterior \textit{pdf}'s for four stars in our sample (their \textit{Gaia} EDR3 \texttt{source\_id} is displayed on top of each panel). 
    The thick black solid line represents $p(d~|~ \Vec{o}, m_{K_s} \Theta_{\mathrm{KM}}, \Theta_{\mathrm{SM}}, \Theta_{\mathrm{CMD}})$. The coloured thin lines illustrate the different components of the \textit{pdf}'s. The correlation terms in the covariance matrix of Eq. A.12 are neglected here for illustrative purposes. The distributions are scaled so that they peak at unity. }
    \label{fig:pdfs}
\end{figure*}

Figure \ref{fig:pdfs} shows example \textit{pdf}'s for four randomly selected stars.
The different components of the \textit{pdf}'s are represented with thin coloured lines, and the total \textit{pdf}, $P(d~|...)$, is represented with a black solid line.
Both the parallaxes of the stars in the top row have $\sigma_{\varpi}/\varpi < 20\%$. In the left panel, the distance estimate is mostly constrained by the term  $P(d~|~\mu_l*, \Theta_{KM}$), which represent the probability of observing a star at distance $d$ given the observed proper motion along Galactic latitude and the kinematic model $\Theta_{KM}$. $P(d~|~\mu_l*, \Theta_{KM}$) shows bi-modality, with a primary maximum at $\sim 6$ kpc, and a secondary maximum at $\sim 1.8$ kpc. The total distance \textit{pdf} shows traces of such bi-modality. In the right panel of the top row instead, the parallax component dominates the distance determination.
The star in the left panel of the bottom row has $\sigma_{\varpi}/\varpi < 20\%$, however both $P(d~|~\mu_l*, \Theta_{KM}$ and $P(d~|~\mu_b, \Theta_{KM}$ peak at closer distances than the parallax term, and thus put a strong constraint on the final distance estimate. Finally, the star in the right panel of the bottom row has $\sigma_{\varpi}/\varpi \approx 70\%$. The final distance \textit{pdf} is however still quite narrow, as a result of the combination of the different terms.

\subsection{Kinematic model}
We assume that stars in our sample follow the rotation curve determined by \cite{Eilers2019}, \begin{equation}v_c = 229 \, \mathrm{km ~s^{-1}}  - 1.7 \mathrm{km ~ s^{-1} ~kpc^{-1}} (R - R_{\odot}),\end{equation} with $R$ the Galactocentric radius and $R_{\odot} = 8.122 \, \mathrm{kpc}$ \citep{Gravity2018}.
The intrinsic velocity dispersions at the Sun's location in the components of the velocity in Galactic coordinates  $U, V, W$ are $\sigma_U = 16.7 $ km/s, $\sigma_V = 10.8 $ km/s, and $\sigma_W = 6$ km/s (Robin et al. 2003). These values are assumed to be constant over the disc.


In Cartesian Galactic coordinates, the \textit{pdf} of the space velocity for one star in our sample is:  
\begin{equation}
    p(\Vec{v} ~|~ \Theta_{KM}) = \frac{1}{(2\pi)^{3/2} |S|^{1/2}} \exp{\left(-\frac{1}{2} (\Vec{v} - \Vec{u})' S^{-1} (\Vec{v} - \Vec{u}) \right)},
\end{equation}
the prime signifying the transpose of the vector.
The quantity $\Vec{u}$ takes into account Galactic rotation and Solar motion, and it can be written as:
\begin{equation}
\vec{u} =\vec{v}_c(\vec{x}_{star}) - \vec{v}_{c}(\vec{x}_{\odot})-\vec{v}_{\odot},
\label{eq.heliovelocity}
\end{equation}
where  
\begin{equation}
\vec{v}_c(\vec{x}_{star}) = v_c(r_{star}) \cdot \bigl ( \hat{\vec{x}}_{star} \times \vec{e}_Z \bigr )
\end{equation}
is the circular velocity in Galactocentric Cartesian coordinates of a star at a Galactocentric distance  $r_{star}$ and $\vec{x}_{star} = [x_{star}, y_{star}, z_{star}]$ .  The unit vectors $\vec{\hat{x}}_{star}$ and $\vec{e}_Z$ are defined as  $\vec{\hat{x}}_{star} = \vec{x}_{star}/|\vec{x}_{star}|$ and $\vec{e}_Z = [0,0,1]$. 
At the Sun's position $\vec{x}_{\odot}$, 
\begin{equation}
\vec{v}_{c}(\vec{x}_{\odot}) = v_c(r_{\odot}) \vec{e}_Y,
\end{equation}
where $\vec{e}_Y = [0,1,0]$ is the unit vector in the $Y$ direction in Galactic coordinates.
We assumed the Sun's position with respect to the Galactic centre to be $\vec{x_{\odot}} = [-8.122, 0., 0.025]$ kpc and the  Sun's peculiar motion with respect to the Local Standard of Rest to be $\vec{v_{\odot}} = [11.1 , 12.24, 7.25]$ km/s \citep{Schonrich2010}.
We write the velocity dispersion matrix $S$ as:
\begin{equation}
\begin{pmatrix}
\sigma_{U}^2 & 0 & 0 \\
0  & \sigma_{V}^2 & 0 \\
0 & 0 & \sigma_{W}^2 
\end{pmatrix}
\end{equation}

For each star the astrometric observables are the parallax $\varpi$, and the proper motions components in $l$ and $b$,  $\mu_{l*} = \mu_l \cos{b}$, and $\mu_b$ respectively. These are collected in arrays: 
\begin{equation}
\Vec{o } = 
\begin{pmatrix}
\varpi - \varpi_0 \\
\mu_{l*}  \\
\mu_b
\end{pmatrix}
\end{equation}
where $\varpi_0$ is the systematic zero point offset of \textit{Gaia} EDR3 parallaxes. There is not a unique  $\varpi_0$ for all the \textit{Gaia} EDR3 stars. 
Following \cite{Bailer-Jones2020}, we applied the parallax zero-point correction derived by \cite{Lindegren2020b}.

By assuming Gaussian errors, the \textit{pdf} for the observables $\Vec{o}$ given the true values 
$\Tilde{\Vec{o}}$, is then:
\begin{equation}
    p(\Vec{o} ~|~ \Tilde{\Vec{o}}) = \frac{1}{(2\pi)^{3/2} |C|^{1/2}} \exp{\left(-\frac{1}{2} (\Vec{o} - \Tilde{\Vec{o}})' C^{-1} (\Vec{o} - \Tilde{\Vec{o}}) \right)}.
\end{equation}
The true values are defined as:
\begin{equation}
\Tilde{\Vec{o }} = 
\begin{pmatrix}
1/d  \\
\Vec{p'} \cdot \Vec{v} /d\, A \\
\Vec{q'} \cdot \Vec{v} /d\, A
\end{pmatrix}
\end{equation}
where $d$ is the true distance to the star and $A  = 4.74047 \, \mathrm{km \, yr/s}$. The two vectors $\Vec{p}$ and $\Vec{q}$ are the two components of the normal triad in longitude and latitude, and, as above, the prime signify their transpose.
The elements of the astrometric covariance matrix $C$ are:\\
\begin{equation}
\begin{pmatrix}
\sigma_{\varpi}^2 & \rho_{\varpi \mu_{l*}} \sigma_{\varpi} \sigma_{\mu_l*} & \rho_{\varpi\mu_{b}} \sigma_{\varpi} \sigma_{\mu_b} \\
\rho_{\varpi\mu_{l*}} \sigma_{\varpi} \sigma_{\mu_l*} & \sigma_{\mu_l*}^2 & \rho_{\mu_{l*}\mu_{b}} \sigma_{\mu_l*} \sigma_{\mu_b} \\
\rho_{\varpi\mu_b} \sigma_{\varpi} \sigma_{\mu_b} & \rho_{\mu_l*\mu_b} \sigma_{\mu_l*} \sigma_{\mu_b} & \sigma_{\mu_b}^2 
\end{pmatrix}
\end{equation}
and are provided in the \textit{Gaia} archive.

The joint \textit{pdf} of the observables with the velocity is therefore:
\begin{equation}
    p(\Vec{o}, \Vec{v} ~|~ d) = p(\Vec{o} ~|~ \Vec{\Tilde{o}}(\Vec{v}, d)) \,  p(\Vec{v} ~|~ \Theta_{KM}).
\end{equation}
The \textit{pdf} for the observables is obtained by marginalising over the velocity:
\begin{equation}
    p(\Vec{o} ~|~ \Theta_{KM}) = \int_{\infty}^{+\infty} d^3\Vec{v} p(\Vec{o} ~|~ \Vec{\Tilde{o}}(\Vec{v}, d)) p(\Vec{v} ~|~ \Theta_{KM}).
\end{equation}
The integral can be resolved analytically. Since the product of two normal \textit{pdf} is normal and the marginal density of a normal \textit{pdf} is also normal, we can write:
\begin{equation}
p(\Vec{o} ~|~ \Theta_{KM}) = \frac{1}{(2\pi)^{3/2} |D|^{1/2}} \exp{\left(-\frac{1}{2} (\Vec{o} - \Vec{c})' D^{-1} (\Vec{o} - \Vec{c}) \right)},
\end{equation}
where 
\begin{equation}
\Vec{c} = 
\begin{pmatrix}
1/d  \\
\Vec{p}' \cdot \Vec{u} /d\, A  \\
\Vec{q}' \cdot \Vec{u} /d\, A
\end{pmatrix}
\end{equation}
and
\begin{equation}
D = C + 
\begin{pmatrix}
0 & 0 & 0  \\
0 & \Vec{p}' S \Vec{p}  & \Vec{p}' S \Vec{q}   \\
0 & \Vec{q}' S \Vec{p}  & \Vec{q}' S \Vec{q}
\end{pmatrix}.
\end{equation}

\subsection{Structural model}
The probability of a star to be at a true distance $d$ is proportional to the stellar density $\rho$ predicted by our spatial model $\Theta_{\mathrm{SM}}$, so that we can write: 
\begin{equation}
    p(d|l,b,\Theta_{\mathrm{SM}})\propto d^2 \rho(d|l,b,\Theta_{SM}),
\end{equation}
where the Jacobian term $d^2$ takes volume effects into account. For convenience we write our model in  Galactocentric coordinates $(R,\phi,z)$, where the stellar density is modelled as an exponential disc: 
\begin{equation}
\rho(R,\phi,z) =\rho_0 \exp \left( -\frac{R-R_{\odot}}{L} \right) \times \exp\left(-\frac{|z|}{h_z} \right)
\end{equation}
$R $ and $z$ are the Galactocentric radius and height above the plane. The quantity $h_z = 0.15 \,  \mathrm{kpc}$ is the disc scale height \citep[cfr.][]{Poggio2020}, and  $L = 3.5 \, \mathrm{kpc}$ is the best fit scale-length of the young ($< 1 $Gyr) disk, where we have used the model presented in \cite{Frankel2018} with the data and method in \cite{Frankel2019}.

\subsection{Luminosity function}
\begin{figure}
    \centering
    \includegraphics[width = \hsize]{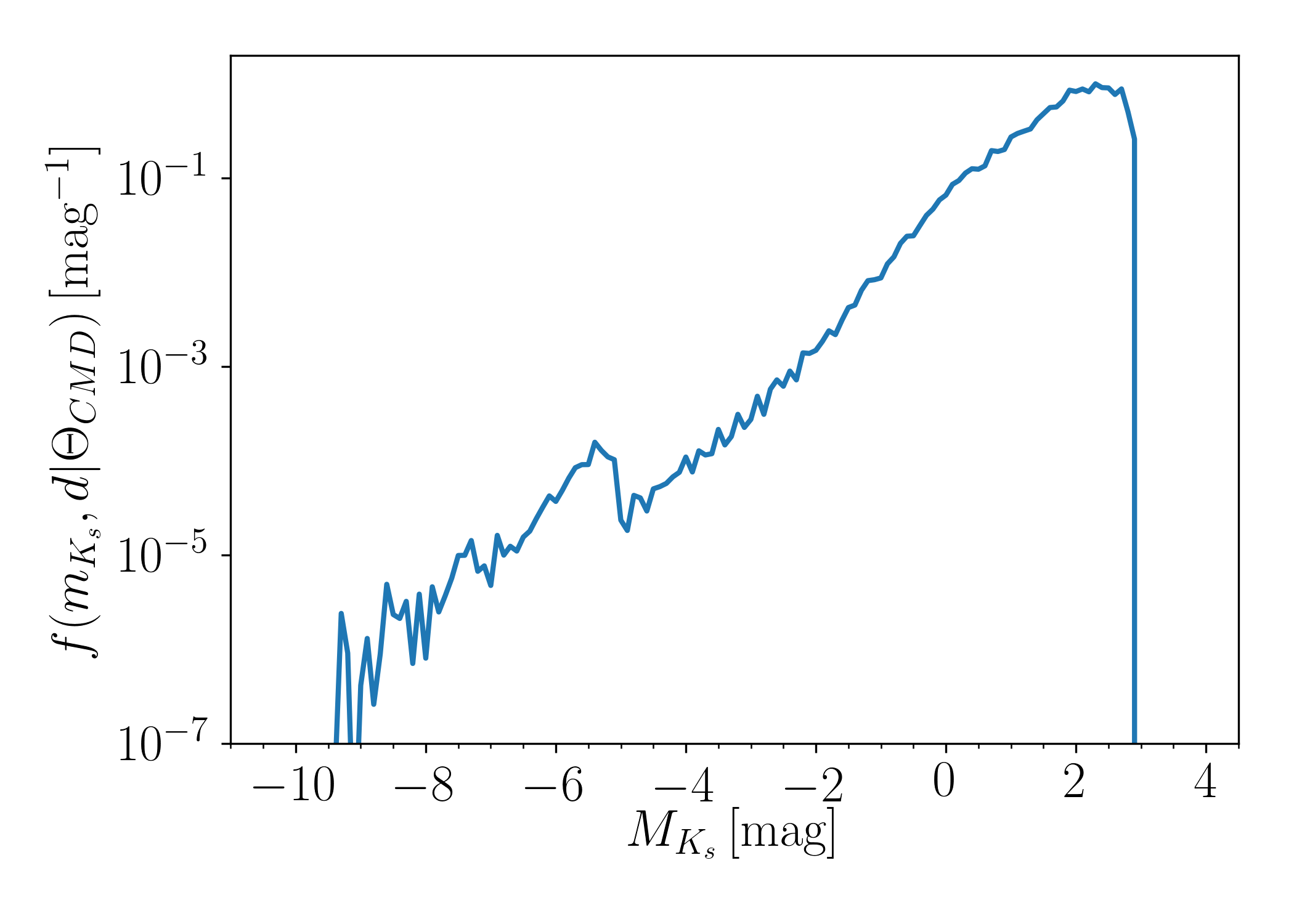}
    \caption{The effective luminosity function $f(d, m_{K_s} | \Theta_{CMD})$ used to determine our prior, normalised to unity.}
    \label{fig:lum}
\end{figure}

We assume that the Milky Way has a universal luminosity function $\phi(M_{K_s}) =  \phi(M_{K_s}(m_{K_s}, d))$, where $M_{K_s}$ is the absolute magnitude in the 2MASS $K_s$ band, and $M_{K_s} = m_{K_s} - 5\log_{10}(d) + 5$. Contrary to \cite{Poggio2020} or \cite{Astraatmadja2016}, we use the near-infrared $K_s$ band to be able to neglect extinction, at least as a first approximation. Following the procedure outlined by \cite{Poggio2020}, we construct the distribution of stars in colour- magnitude space $\phi(m_{K_s}, d, c)$ by assuming a constant star formation rate, the two-part power law Kroupa initial mass function corrected for unresolved binaries, and solar metallicity.  The term $c$ indicates a colour (for instance, $G - K_s$). The distribution of stars in colour-magnitude space is then modified by setting to zero the parts of the distribution where $T_{\mathrm{eff}} < 8000 \mathrm{K}$. This roughly corresponds to applying the colour cuts aiming at selecting hot stars as described in Sec. \ref{sec:selection}, and effectively changes the term $S(c)$ of Eq. \ref{eq:effective_LF} into a selection approximately based on effective temperature, $S(T_\mathrm{eff})$. We can thus write the \textit{effective} luminosity function, shown in Fig. \ref{fig:lum}, $f(d, m_{K_s} ~|~ \Theta_{\mathrm{CMD}})$ as:
\begin{equation}\label{eq:effective_LF}
    f(d, m_{K_s} ~|~ \Theta_{\mathrm{CMD}}) = \int \phi(m_{K_s}, d, c) S(c) dc,
\end{equation}
where $S(c) = 1$ if $T_{\mathrm{eff}} > 8000$ K and  $S(c) = 0$ otherwise.

\subsection{Distance uncertainties}\label{sec:distance_uncertainties}
\begin{figure}
    \centering
    \includegraphics[width = \hsize]{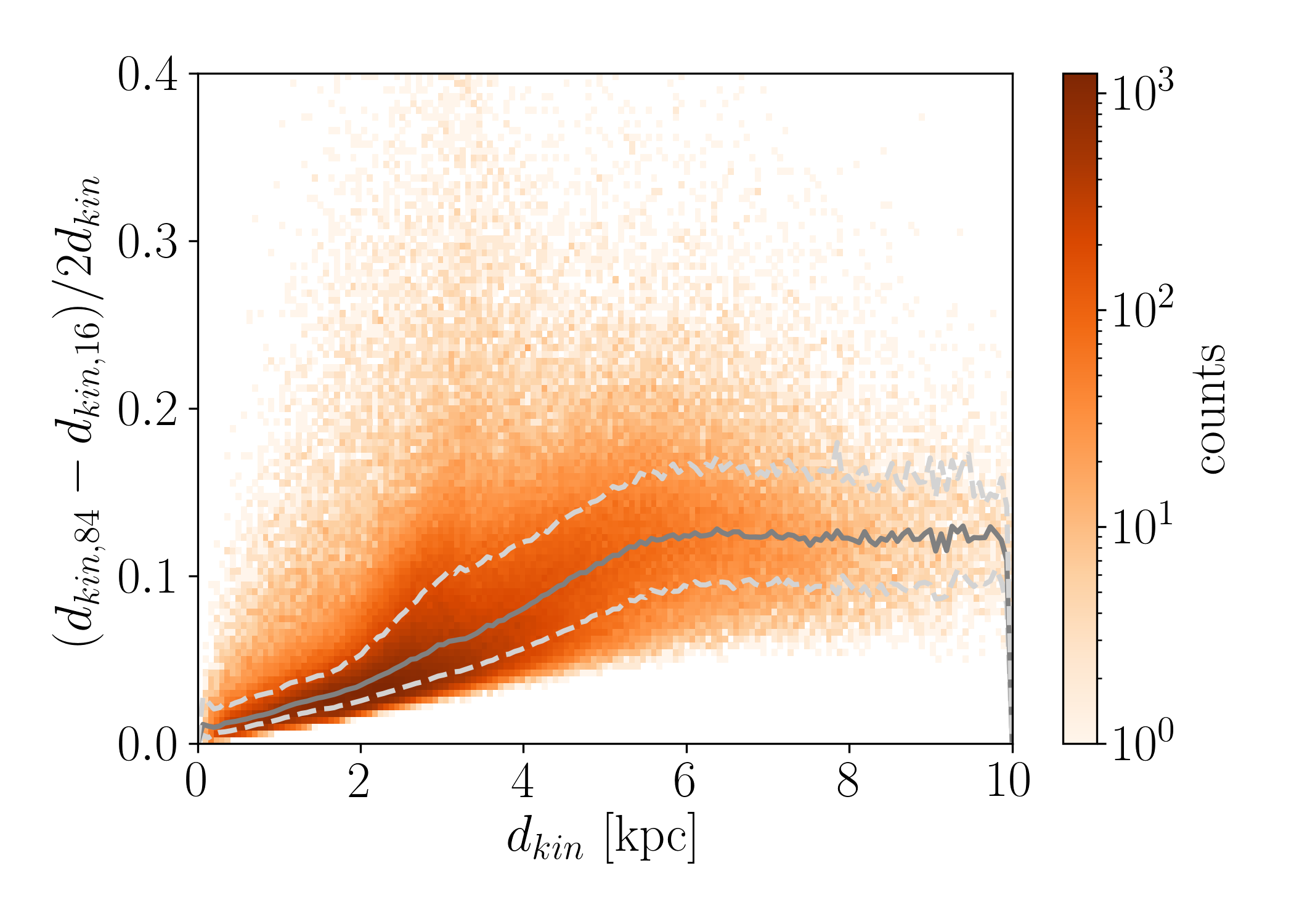}
    \caption{Fractional distance uncertainty $d_{kin, 84} - d_{kin, 16}/ 2~d_{kin}$ for stars in the clean sample. The dashed light gray lines correspond to the 16$th$ and 84$th$ percentiles, the solid gray line corresponds to the median fractional error. }
    \label{fig:frac_uncertainty}
\end{figure}

Figure \ref{fig:frac_uncertainty} shows the fractional distance uncertainty for stars in the filtered sample (see Section \ref{sec:filtered_sample}). For distances smaller than $\sim$4~kpc, parallax uncertainties dominate the fractional error distribution. This is a regime where  $\sigma_{\varpi}/\varpi < 0.1$, and the inverse parallax is a  good distance estimate \citep[see also][]{Bailer-Jones2020}. For distances larger than $\sim 4$~kpc, the kinematic model and the other priors described above keep the distance uncertainties from exploding. The fractional errors
are below the 10\% for $d_{kin} < 5$~kpc, and remain well below the 20\% for $d_{kin} < 10$~kpc.

\subsection{Distance validation}
\begin{figure}
    \centering
    \includegraphics[width = \hsize]{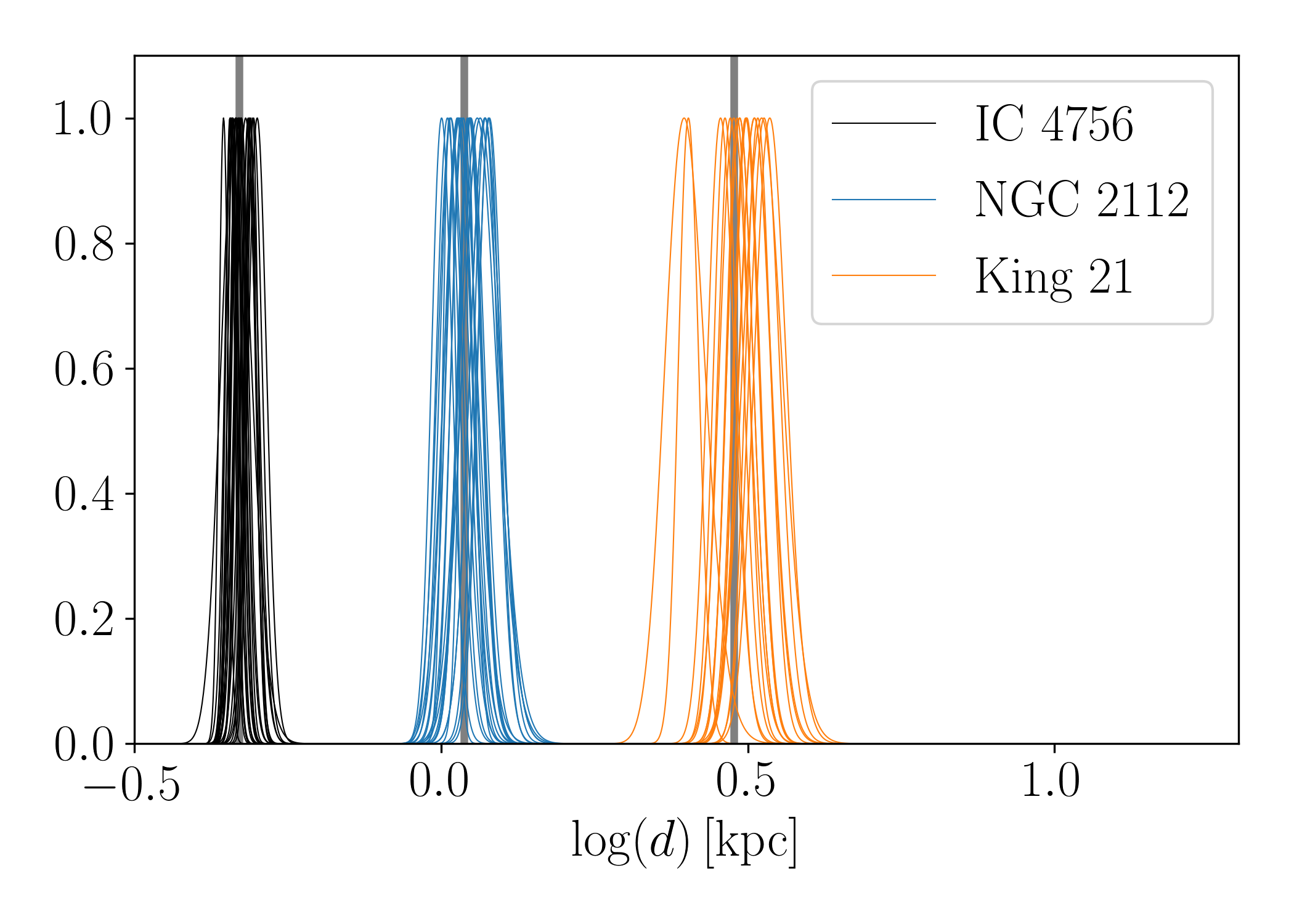}
    \caption{Single star distance $pdf$'s for the members of the three clusters IC 4756 (black), NGC 2112 (blue), and King 21 (orange). The grey vertical lines represent the cluster distances (respectively 469pc, 1090pc, and 3005pc) derived by \cite{Cantat2020}.}
    \label{fig:cluster_pdfs}
\end{figure}

To validate our distances, we identified the stars in our sample that belong to the three clusters IC 4756, NGC 2112, and King 21, respectively at 469 pc, 1090 pc, and 3005 pc, as determined  by \cite{Cantat2020}.
Fig. \ref{fig:cluster_pdfs} shows the distance $pdf$'s of such stars. As exepcted, the $pdf$'s for stars at larger distances are broader than for those at closer distances. The distance estimates for all the cluster members are compatible within 2$\sigma$ with the average distance value. The dispersion around the median distance is between 10-20 pc for IC 4756, around 65 pc for NGC 2112, and around 300 pc for King 21. This indicates that distance uncertainties are within the 10\% for  $d \sim 3$ kpc (where most of the over-densities are located), in agreement with Sec \ref{sec:distance_uncertainties}. 

\section{Vertical velocity}\label{sec:vertical velocity}
As described in \cite{Drimmel2000a},  the vertical velocity $v_{z}$ of a star is computed by:
\begin{equation}
    v_z = 4.74047 \, \mu_b\, d /\cos(b) + W_{\odot} + v_r \sin(b),
\end{equation}
where $b$ is Galactic longitude, $\mu_b$ is the proper motion along $b$, $d$ is the distance estimate, $W_{\odot}$ is the Sun's velocity along $Z$, and $v_r$ is the line-of-sight velocity.
The majority of stars in our sample lacks line-of-sight velocities, thus it is not possible to calculate 
directly their vertical velocity. However, for stars at low Galactic latitudes ($|b| < 20^{\circ}$), the term $v_r \sin(b)$ can be approximately computed by assigning to each star the velocity it would have if it followed exactly the Galactic rotation curve:
\begin{equation}
     v_r \sin(b) \approx (S - S_{\odot}) \tan(b), 
\end{equation}
where: $S_{\odot} = U_{\odot} \cos{l} + V_{\odot}\sin(l)$, with $l$ the galactic longitude of a star, and $U_{\odot}$ and $V_{\odot}$ the components of the Sun's motion along $X$ and $Y$ respectively;
$S = v_{\phi} \, R_{\odot}/ R - v_{LSR} \sin (l)$, with $R_{\odot}$ the distance of the Sun from the Galactic Centre, $R$ a star's Galactocentric radius, $v_{\phi}$ the azymutal velocity, and $v_{LSR}$ the standard of rest velocity.
We assumed $(U, V, W)_{\odot} = (11.1, 12.24, 7.25)$ km/s from  \cite{Schonrich2010}, and the rotation curve derived by \cite{Eilers2019}, reported in Eq. B.3.

%
%

\end{document}